\documentclass[showpacs,aps,floatfix]{revtex4}
\usepackage{graphicx}
\usepackage{amsmath}
\usepackage{bm}
\begin{document}
\title{Investigation of Phase Diagram and Bulk Thermodynamic 
Properties using PNJL Model with Eight-Quark Interactions}
\author{Abhijit Bhattacharyya}
\email{abphy@caluniv.ac.in}
\author{Paramita Deb}
\email{paramita.deb83@gmail.com}
\affiliation{Department of Physics, University of Calcutta,
92, A. P. C. Road, Kolkata - 700009, INDIA}
\author{Sanjay K. Ghosh}
\email{sanjay@bosemain.boseinst.ac.in}
\author{Rajarshi Ray}
\email{rajarshi@bosemain.boseinst.ac.in}
\affiliation{Department of Physics and Centre for Astroparticle Physics \&
Space Science, Bose Institute,
93/1, A. P. C Road, Kolkata - 700009, INDIA}

\vskip 0.3in
\begin {abstract}
 We present the bulk thermodynamic properties and phase diagram of 
strongly interacting matter in an extension of the 3-flavor 
NJL and PNJL models of QCD. Using a three momentum cut-off scheme, 
we have extended the multiquark interaction terms up to eight order
so that the stability of the vacuum is ensured in these models.
We explore the effects of various combinations of the two 
eight-quark couplings $g_1$ and $g_2$ and present a comparative study
between the NJL and PNJL models as well as Lattice QCD data.
The main effect of the eight-quark interaction term is to shift
the critical end point in the $T-\mu$ phase diagram to a lower
value of $\mu$ and higher value of $T$, thus bringing them closer 
to Lattice QCD results. 
\end{abstract}

\maketitle

\vskip 0.3in
{\section {Introduction}}
The strong interaction as described by Quantum Chromodynamics (QCD), 
is a remarkable branch of physics which shows a rich phase structure 
at finite temperature and density. At low temperatures and densities, 
the dominant 
degrees of freedom in our nature are color-singlet bound states of 
hadrons. However, due to the asymptotic freedom of QCD, it is expected 
that at very high temperatures and densities these hadrons break up to 
liberate quarks and gluons and form the quark-gluon plasma (QGP). The 
experimental exploration of such a phase transition from the confined 
hadronic phase to the deconfined QGP phase is being pursued actively
in the Relativistic Heavy Ion Collider (RHIC) and more data are 
expected from Large Hadron Collider (LHC) running at CERN and in the 
future experiment at FAIR.

 The transition region is however quite far from the asymptotic regime
of QCD. This non-perturbative nature makes the study of these hot 
and dense matter quite non-trivial. The most reliable way to analyze the
physics in this range of interest is to perform the numerical 
computation of a lattice version of the color SU(3) gauge theory 
(Lattice QCD/LQCD). The scheme is robust but numerically costly. Hence
the most popular way to study the physics of the phase transition of 
strongly interacting matter is to look into one of the
various possible effective models inspired from QCD. One of these is 
the Nambu-Jona-Lasinio (NJL) model \cite{Nambu,klev,hat1,hat2} which
incorporates the global symmetries of QCD
quite nicely. A four quark interaction term in the NJL Lagrangian is 
able to generate the physics of spontaneous breaking of chiral symmetry 
- a property of QCD which is manifested as the nondegenerate chiral 
partners of the low-mass hadrons.
But the major drawback of the NJL model is to have a reasonable 
description of the physics of color confinement. In this respect, 
the Polyakov loop extended NJL (PNJL) model 
\cite{fuku,ratti,ray,gatto} tries to incorporate the fact that 
the chiral transition in QCD is of deconfining nature by introducing 
a background temporal gluon field.

In our previous work \cite{deb} we elaborately studied the 2+1 
flavor PNJL model \cite{gatto}, incorporating four-quark and six-quark 
interaction terms in the Lagrangian with three-momentum
cut-off regularization. The six-quark interaction term (or the 't Hooft 
determinant term) preserves the ${SU(3)}_L\times{SU(3)}_R$ symmetry 
and breaks the ${U(1)}_A$ symmetry required due to the 
chiral anomaly. This term is also responsible for the flavor mixing 
in the $\eta_0$ and $\eta_8$ mesons to give $\eta$ and $\eta^\prime$
mesons. 

  However, due to the six-quark term the vacuum of the model becomes unbounded 
from below. This problem has become more obvious in the functional 
analysis of the Lagrangian. In the work of Reinhardt and Alkofer 
\cite{alkofer} the stationary phase approximation method was 
considered to bosonize the model and to calculate the contribution of
the classical path at the lowest order. This lowest order results sums
all tree diagrams in powers of the coupling constant of the six quark
interaction. But the functional integral bosonization of the model 
shows several classical trajectories in that interval of the 
functional integration. Since at leading order there can be only one 
classical trajectory in the mean-field potential, the semiclassical
potential derived from the functional integration differs from the 
mean-field one. So if we consider several classical trajectories, the 
effective potential of the theory gets unbounded from below. 
 
  In fact, the origin of multi-quark interactions is not very 
well-known. But we can find evidences of multi-quark interactions in 
the semi-classical theories based on QCD instanton vacuum \cite{shuryak}.
There we can find certain correlations between two or more quarks by 
averaging over their positions and orientations in color space. Also in 
the instanton-gas model an infinite number of multi-quark interactions 
can be found beyond the zero mode approximation \cite{yu}. There are 
also some lattice measurements for the QCD vacuum which shows a 
hierarchy between the multi-quark interaction \cite{yu1}. In this case 
the lowest four quark interaction term forms a stable vacuum
by breaking the chiral symmetry spontaneously. But the next term in the
hierarchy, the six quark interaction term, which is needed to mimic
the $U_A(1)$ anomaly, destroys the ground state. So 
one cannot truncate the tower of multi-quark interactions at this level. 
The next candidate is the eight-quark interaction term which ensures the
stability of the vacuum. 

  There are some recent findings \cite{osipov1,osipov2,hiller} which show that
the addition of the eight-quark interaction term in the Lagrangian may 
solve the problem of unstable vacuum. The (2+1)-flavor NJL model with 
an eight-quark interaction was studied in the four momentum cut-off
and Pauli-Villars regularization scheme. The standard bosonization 
procedure of six and eight-quark interactions was followed and the 
multi-quark vertices were replaced by purely mesonic ones by the 
stationary-phase method. Indeed the controversy between the mean-field 
approach and the functional integral approach has been removed by 
including the eight-quark interaction term in the Lagrangian, since it 
restricts the number of classical trajectories to one, giving a stable 
ground state of the system. Recently, Kashiwa {\it et.al.} 
\cite{kashiwa1,kashiwa2} have studied the chiral phase transition in 
the 2-flavor NJL and PNJL models with eight-quark interactions. The
2-flavor model does not incorporate the six-quark interaction.
 
 In this paper we first of all extend the (2+1)-flavor NJL and PNJL
models to include the eight-quark interactions at non-zero temperatures
and densities. For this we use the 3-momentum cut-off regularization
scheme. Thereafter we thoroughly investigate the effect of the 
eight-quark term in the thermodynamic properties of strongly interacting
matter.

  We organize our paper as follows: in the next section we briefly 
describe the 2+1 flavor PNJL model with eight-quark interaction term 
in the Lagrangian. For the Polyakov loop potential we incorporate
Vandermonde term \cite{ray}.
In sec. ${III}$ we describe the thermodynamic potential
of the PNJL model. We elaborately discuss the parameter sets chosen,
and the stability criteria. The next two sections discuss the
results of the paper, about the chiral phase transition and the bulk
thermodynamic properties like pressure, trace anomaly at zero density and 
quark number density at finite temperature and density. We give a detailed
analysis of the phase diagram and the critical end points at the next section.
A summary of our results are available in the last section.

\vskip 0.3in
{\section {  Three Flavor PNJL model with eight-quark interaction }}

In this section we briefly describe the formalism of the PNJL model with
eight-quark interaction term. Some details on three flavor PNJL model with 
four and six quark interaction can be found in the literature 
\cite{ratti,fuku,gatto,deb} and the NJL model with eight-quark interaction 
term in \cite{osipov1,osipov2,hiller}. There is also some work on two 
flavor PNJL model with eight-quark interaction term \cite{kashiwa2}. 

 In the PNJL model the gluon dynamics is reduced to the chiral point 
couplings between quarks and a simple temporal background gauge field 
which represents Polyakov Loop dynamics. The Polyakov line is represented 
as (see e.g. \cite{dumi} and references therein),
\begin {equation}
  L(\bar x)={\cal P} {\rm exp}[i {\int_0}^\beta d\tau A_4{({\bar x},\tau)}]
\end {equation}
where $A_4=iA_0$ is the temporal component of Eucledian gauge field
$(\bar A,A_4)$, $\beta=\frac {1}{T} $, and $\cal P$ denotes path ordering.
$L(\bar x)$ transforms as a field with charge one under global Z(3) 
symmetry. The gluon dynamics can be described as an effective theory 
of the Polyakov loops \cite{pisarski1}. The quark thermodynamics can be 
described by the NJL model \cite{Nambu}. Let us consider the 
$SU(3)_f$ version of PNJL model with eight-fermion interaction described 
by the Lagrangian,  
\begin {align}
   {\cal L} &= {\sum_{f=u,d,s}}{\bar\psi_f}\gamma_\mu iD^\mu
             {\psi_f}-\sum_f m_{f}{\bar\psi_f}{\psi_f}
              +\sum_f \mu \gamma_0{\bar \psi_f}{\psi_f}\nonumber\\
       &+{\frac {g_S} {2}} {\sum_{a=0,\ldots,8}}[({\bar\psi} \lambda^a {\psi})^2+
            ({\bar\psi} i\gamma_5\lambda^a {\psi})^2]
       -{g_D} [det{\bar\psi_f}{P_L}{\psi_{f^\prime}}+det{\bar\psi_f}
            {P_R}{\psi_{f^\prime}}]\nonumber\\
  &+8{g_1}[({\bar\psi_i}{P_R}{\psi_m})({\bar\psi_m}{P_L}{\psi_i})]^2
    +16{g_2}[({\bar\psi_i}{P_R}{\psi_m})({\bar\psi_m}{P_L}{\psi_j})
          ({\bar\psi_j}{P_R}{\psi_k})({\bar\psi_k}{P_L}{\psi_i})]\nonumber\\ 
        &-{\cal {U^\prime}}(\Phi[A],\bar \Phi[A],T)\nonumber\\
  &= {\cal L}_0+{\cal L_{SB}}+{\cal L}_\mu+{\cal L}_s+{\cal L_{KMT}}+
       {\cal L}_{8q}^1+{\cal L}_{8q}^2-{\cal {U^\prime}}
\label{lag}
\end {align}
where the matrices $P_{L,R}=(1\pm \gamma_5)/2$ are chiral projectors.
In the above Lagrangian ${\cal L}_0$ is the Dirac term with gauge field
interactions; $D^\mu=\partial^\mu-i{A_4}\delta_{\mu 4}$. 
${\cal L}_{SB}$ is the
mass term which breaks the symmetry explicitly. The mass of a particular
flavor is denoted by $m_f$, where $f=u,d,s$. 
${\cal L}_s$ is the four-fermi 
interaction term with coupling $g_S$. 
The next term, ${\cal L}_{KMT}$, which is a six-fermi interaction term, is
invariant under ${SU(3)}_L\times {SU(3)}_R$ but breaks $U(1)_A$ symmetry.
This term mimics the QCD chiral anomaly. The terms ${\cal L}_{8q}^1$ and 
${\cal L}_{8q}^2$ are the eight-quark interaction terms which describe 
the spin zero interactions where $g_1$ and $g_2$ are the corresponding 
couplings. Here we have considered the interaction terms to be 
effectively local since the meson physics in the large $N_c$ limit is 
described by the local Lagrangian of this type. Since the coupling 
constants are dimensionful, the model is not renormalizable. So we have 
used three-momentum cut-off regulator $\Lambda$ to make quark loops
finite. 

  In earlier versions of the PNJL model we found that the Polyakov loop,
which is the normalized trace of the Wilson line $L$, has become 
greater than $1$ above $2T_C$ \cite{ray1,mukherjee,deb}.
To solve this problem one has to take a proper Jacobian of 
transformation from the matrix valued field $L$ to $\Phi$ which will then
constrain the value of $\Phi$ within 1. One way to resolve this 
problem is to introduce Vandermonde term in the Polyakov loop potential
\cite{ray}. Thus the potential $\cal {U^\prime}$ with the Vandermonde 
term can be expressed as 
\begin{equation}
\frac {{\cal {U^\prime}}(\Phi[A],\bar \Phi[A],T)} {T^4}= 
\frac  {{\cal U}(\Phi[A],\bar \Phi[A],T)}{ {T^4}}-\kappa \ln[J(\Phi,{\bar \Phi})]
\label {uprime}
\end{equation}
where $\cal {U(\phi)}$ is the Landau-Ginzburg type potential given by
 \cite{ratti},
\begin{equation}
\frac  {{\cal U}(\Phi, \bar \Phi, T)}{  {T^4}}=-\frac {{b_2}(T)}{ 2}
                 {\bar \Phi}\Phi-\frac {b_3}{ 6}(\Phi^3 + \bar \Phi^3)
                 +\frac {b_4}{  4}{(\bar\Phi \Phi)}^2
\end{equation}
with
\begin {eqnarray}
\Phi = (Tr_c L)/N_c, {\hspace{0.3in}} {\bar \Phi} = (Tr_c L^\dagger)/N_c
\nonumber \\
     {b_2}(T)=a_0+{a_1}(\frac { {T_0}}{ T})+{a_2}(\frac {{T_0}}{ T})^2+
              {a_3}(\frac {{T_0}}{T})^3,
\end {eqnarray}
$b_3$ and $b_4$ being constants. The second term in eqn. \ref{uprime} is 
known as Vandermonde term, where $J(\Phi, {\bar \Phi})$ is the Jacobian of 
transformation from Wilson line $L$ to $(\Phi, {\bar \Phi})$ written as
\begin {equation}
J[\Phi, {\bar \Phi}]=(27/24{\pi^2})(1-6\Phi {\bar \Phi}+\nonumber\\
4(\Phi^3+{\bar \Phi}^3)-3{(\Phi {\bar \Phi})}^2
\end{equation}
$J(\Phi, {\bar \Phi})$ is also known as Vandermonde determinant and is not
explicitly space time dependent. The 
value of the dimensionless parameter $\kappa$ will be determined 
phenomenologically. The coefficient $\kappa$ in the VDM term can in general
have some temperature and/or chemical potential dependence. But here we take
a constant value in such a way that we
can get the pressure and the transition temperature as close as possible 
to the lattice QCD results for the PNJL model with and without the 
eight-quark interaction. This coefficient has been tabulated later.   
  
 In order to study the chiral transition of the system we consider 
the Mean Field Approximation (MFA) of the eqn. (\ref{lag}) to get the 
field equations for $\Phi$, $\bar \Phi$, $\sigma$, where it is assumed 
that the system is described as an assembly of non-interacting particles
moving in the mean field. The theory is analogous to the BCS theory of
superconductor, where the pairing of two electrons leads to the 
condensation causing the gap in the energy spectra. Similarly in this 
model, due to the dynamical breaking of ${SU(3)}_L \times {SU(3)_R}$ 
symmetry to $SU(3)_V$ a composite operator picks up a nonzero vacuum 
expectation value leading to $\langle{\bar \psi}\psi\rangle$ condensation.
Due to the dynamical breaking of chiral symmetry, nine goldstone bosons 
appear for $N_f=3$ model. The quark condensate is given as,
\begin {equation}
 \langle{\bar \psi_f}{\psi_f}\rangle= 
-i{N_c}{{{\cal L}t}_{y\rightarrow x^+}}(tr {S_f}(x-y))
\end {equation}
where trace is over color and spin states. The self-consistent gap 
equation for the constituent masses are,
\begin {equation}
  M_f =m_f-2g_S \sigma_f+\frac {{g_D}}{2} \sigma_{f+1}\sigma_{f+2}-2g_1 
     \sigma_f{(\sigma_u^2+\sigma_d^2+\sigma_s^2)}-4g_2 \sigma_f^3  
\end {equation}
here $\sigma_f=\langle{\bar \psi_f} \psi_f\rangle$ denotes chiral 
condensate of the quark with flavor $f$. $f$ denotes the flavor $u,d,s$
respectively. Here if we consider $\sigma_f=\sigma_u$, then 
$\sigma_{f+1}=\sigma_d$ and $\sigma_{f+2}=\sigma_s$. Similarly if 
$\sigma_f=\sigma_d$ then $\sigma_{f+1}=\sigma_s$ and $\sigma_{f+2}=\sigma_d$,
if $\sigma_f=\sigma_s$ then $\sigma_{f+1}=\sigma_u$ and 
$\sigma_{f+2}=\sigma_d$. The expression for $\sigma_f$ at $T=0$ and 
$\mu=0$ can be written as \cite{gatto}
\begin {equation}
 \sigma_f=-\frac {3{M_f}}{ {\pi}^2} {{\int_0}^\Lambda}\frac {p^2}{
           \sqrt {p^2+{M_f}^2}}dp,
\end {equation}
$\Lambda$ being the three-momentum cut off.
\vskip 0.3in

{\section {Thermodynamic Potential and parameter fitting}
\label{prmtfit}}

The primary aim of our work is to study the thermodynamic properties of
the strongly interacting matter using the PNJL model with eight-quark 
interactions at zero and finite chemical potentials. To do so we now 
need to fix the parameters in both the NJL and Polyakov loop potentials. 
The thermodynamic potential for the multi-fermion interaction in MFA of 
the PNJL model can be written as
\begin {align}
  \Omega (\Phi,\bar \Phi,M,T,\mu)&= {\cal {U^\prime}}[\Phi,\bar \Phi,T]+
               2{g_S}{\sum_{f=u,d,s}}
            {\sigma_f^2}-\frac {{g_D}}{ 2}{\sigma_u}
          {\sigma_d}{\sigma_s}+3\frac {{g_1}}{2}({\sigma_f^2})^2
           +3{g_2}{\sigma_f^4}\nonumber\\
                &-T{\sum_n}\int\frac {{d^3p}}{{(2{\pi})^3}}
                {Tr} \ln\frac {{ S^{-1}}(i{\omega_n},\bar p)}{ T}
\end {align}
where $\omega_n=\pi T(2n+1)$ are Matsubara frequencies for fermions.
The inverse quark propagator is given in momentum space by
 \begin {equation}
   {S^{-1}}=\gamma_0(p^0+\mu-i{A_4})-\vec{\gamma} \cdot{\vec p}-M
\end {equation}
using the identity ${Tr}\ln\left(X\right)=\ln \det\left(X\right)$, we get

\begin {align}
 \Omega &= {\cal {U^\prime}}[\Phi,\bar \Phi,T]+2{g_S}{\sum_{f=u,d,s}}
            {\sigma_f^2}-\frac {{g_D} }{2}{\sigma_u}
          {\sigma_d}{\sigma_s}+3\frac {{g_1}}{2}({\sigma_f^2})^2\nonumber\\
           &+3{g_2}{\sigma_f^4}-6{\sum_f}{\int_{0}^{\Lambda}}
     \frac {{d^3p}}{{(2\pi)}^3} E_{pf}\Theta {(\Lambda-{ |\vec p|})}\nonumber \\
       &-2{\sum_f}T{\int_0^\infty}\frac {{d^3p}}{{(2\pi)}^3}
       \ln\left[1+3(\Phi+{\bar \Phi}e^{\frac {-(E_{pf}-\mu)}{ T}})
       e^{\frac {-(E_{pf}-\mu)}{ T}}+e^{\frac {-3(E_{pf}-\mu)}{ T}}\right]
\nonumber\\
       &-2{\sum_f}T{\int_0^\infty}\frac {{d^3p}}{{(2\pi)}^3}
        \ln\left[1+3({\bar \Phi}+{ \Phi}e^{\frac {-(E_{pf}+\mu)}{ T}})
       e^{\frac {-(E_{pf}+\mu)}{ T}}+e^{ \frac {-3(E_{pf}+\mu)}{ T}}\right]
\end {align}
where $E_{pf}=\sqrt {p^2+M^2_f}$ is the single quasiparticle energy,
$\sigma_f^2=(\sigma_u^2+\sigma_d^2+\sigma_s^2)$ and 
$\sigma_f^4=(\sigma_u^4+\sigma_d^4+\sigma_s^4)$.
In the above integrals, the vacuum integral has a cutoff $\Lambda$ 
whereas the medium dependent integrals have been extended to infinity. 
The main idea to incorporate the eight-quark interaction term in the 
Lagrangian is to stabilize the vacuum. The detailed discussion of the 
stability criteria of the vacuum can be found in the literature 
\cite{osipov1,osipov2}.
\begin{table}
\begin{center}
\begin{tabular}{|c|c|c|c|c|c|c|c|c|}
\hline
\multicolumn{8}{|c|}{Sets}&Physical \\
\cline{1-8}
\multicolumn{4}{|c|}{six-quark}&\multicolumn{4}{|c|}{eight-quark}
             &Parameter \\
\cline{1-8}
$a $&$ b $&$ c $&$ d $&$ e $&$ f $&$ g$&$h $& values \\
\hline
$m_\pi$&$m_\pi$&$m_\pi$&$m_\pi$&$m_\pi$&$m_\pi$&$m_\pi$&$m_\pi$&$
m_\pi=138 \, \rm MeV$\\
$f_\pi$&$f_\pi$&$f_\pi$&$f_\pi$&$f_\pi$&$f_\pi$&$f_\pi$&$f_\pi$&$
f_\pi=93 \, \rm MeV$\\
$m_k$&$m_k$&$m_k$&$m_k$&$m_k$&$m_k$&$m_k$&$m_k$&$m_k=494 \, \rm MeV $\\
$-$&$-$&$-$&$-$&$f_k$&$f_k$&$f_k$&$f_k$&$f_k=117 \,  \rm MeV$\\
$m_\eta$&$m_\eta$&$m_\eta$&$-$&$m_\eta$&$m_\eta$&$m_\eta$&$-$&$
m_\eta=480 \, \rm MeV $\\
$-$&$-$&$-$&$-$&$m_{\eta^\prime}$&$m_{\eta^\prime}$&$m_{\eta^\prime}$&$
m_{\eta^\prime}$&$m_{\eta^\prime}=957 \, \rm MeV$\\
$-$&$m_\sigma$&$-$&$m_\sigma$&$-$&$m_\sigma$&$-$&$m_\sigma$&$m_\sigma=680 \, \rm MeV$\\
$m_u$&$-$&$-$&$m_u$&$m_u$&$-$&$-$&$m_u$&$m_u= 5.5 \,  \rm MeV$\\
$-$&$-$&$\sigma_u$&$-$&$-$&$-$&$\sigma_u$&$-$&$\sigma_u^{1/3}=-248 \,  \rm MeV$\\

\hline
\end{tabular}
\caption{(Color online) The PNJL model parameters are fitted to 
reproduce the different physical parameters as indicated above.}
\label{table1}
\end{center}
                                                                                
\end{table}

The parameters of the NJL part of the Lagrangian are the current quark
masses $m_u$, $m_d$ and $m_s$, the coupling constants $g_s$, $g_D$, 
$g_1$ and $g_2$ and the three-momentum cutoff $\Lambda$ characterizing 
the scale of the chiral symmetry breaking. Here we consider the isospin 
symmetric limit $m_u=m_d$. These parameters are determined by 
reproducing few physical quantities like pion mass $m_\pi=138 \, \rm MeV$,
kaon mass $m_K=494 \, \rm MeV$, eta mass $m_\eta=480 \, \rm MeV$, eta prime
mass $m_\eta^\prime=957 \, \rm MeV$, pion decay constant 
$f_\pi=93 \, \rm MeV$, kaon decay constant $f_K=117 \, \rm MeV$, u quark 
condensate $\sigma_u^{1/3}=-248 \, \rm MeV$ and/or the mass of sigma meson 
$m_\sigma=680 \, \rm MeV$.

\begin{table}
\begin{center}
\begin{tabular}{|c|c|c|c|c|c|c|c|c|c|c|}
\hline
Set &$ m_u $&$ m_s $&$ \Lambda $&$ g_S \Lambda^2 $&$ g_D \Lambda^5 $&$  
g_1 \times 10^{-21}$&$ g_2 \times 10^{-22}$&$ \kappa $&$ 
T_C^{PNJL} $&$ T_C^{NJL}$\\

& MeV & MeV & MeV & & & MeV$^{-8}$ &  MeV$^{-8}$ & & MeV & MeV \\
\hline

$ a $&$ 5.5 $&$ 134.758 $&$ 631.357 $&$ 3.664 $&$ 74.636
$&$ 0.0 $&$ 0.0 $&$ .13 $&$ 181.0 $&$ 170.55$\\

$ b$&$ 5.406 $&$ 133.227 $&$ 641.357 $&$ 
3.717 $&$ 61.309 $&$ 0.0 $&$ 0.0 $&$ .18 $&$ 182.2 $&$170.25$  \\
$ c $&$ 5.418 $&$ 133.562 $&$ 640.206 $&$ 3.637 $&$ 70.849
$&$ 0.0 $&$ 0.0 $&$ .14 $&$ 180.5 $&$169.25 $ \\
$d $&$5.5 $&$ 133.532 $&$ 631.337 $&$
4.229 $&$ 14.461 $&$ 0.0 $&$ 0.0 $&$ .11 $&$ 183.5 $&$ 176.65$\\
                                                                                
$e$ & $ 5.5 $&$ 183.468 $&$ 637.720 $&$ 2.914 $&$ 75.968
$&$ 2.193 $&$ -5.890 $&$ .06 $&$ 168.5 $&$ 141.35$\\

$f $&$12.509 $&$ 181.863 $&$ 628.933 $&$ 
2.986 $&$ 75.444 $&$ 2.007
$&$ -4.538 $&$ .07 $&$ 171.8 $&$ 150.05$\\
$g$ &$ 8.742 $&$ 179.498 $&$ 640.206 $&$
2.928 $&$ 75.382 $&$ 1.929 $&$
-4.840 $&$ .05 $&$ 169.0 $&$ 144.25$\\

                                                                                
 $h$ & $ 5.5 $&$ 187.786 $&$ 628.933 $&$ 2.956 $&$ 75.983
$&$ 2.425 $&$ -6.445 $&$ .08 $&$ 170.5 $&$ 144.15 $\\

\hline
\end{tabular}
\caption{(Color online) Parameters of the $SU(3)$ NJL part and different value of $T_C$ at 
$\mu=0$ and the value of $\kappa$ for all input parameter sets. }
\label{table2}
\end{center}

\end{table}

We consider two cases of up to six-quark ($g_1=0=g_2$) and up to eight-quark
($g_1\ne 0 \ne g_2$) interactions each with four sets $(a,b,c,d)$ and
$(e,f,g,h)$ respectively. In table \ref{table1} we have tabulated 
different physical observables and their values for the different sets. 
In set $(a,d)$ and $(e,h)$  $m_u$ is kept fixed at $5.5 \, \rm MeV$ at 
$1$ $ \rm GeV$ scale since the values determined from the evaluation of 
the current matrix elements at low energies are also centered around 
$ 5.5 \, \rm MeV$. In sets $b$ and $f$ the mass of sigma meson is considered
as fixed to obtain the model parameters whereas in sets $c$ and $g$ the 
u quark condensate is considered. In sets $(b,c)$ we fit the parameters
$(m_u,m_s,\Lambda,g_S,g_D)$ by fixing $(m_\pi,m_K,m_\eta,f_\pi$ and
$m_\sigma$ or ${\sigma_u})$  and in sets $(f,g)$ the parameters
$(m_u,m_s,\Lambda,g_S,g_D,g_1,g_2)$ are obtained by fixing $(m_\pi,
m_K,m_\eta,m_\eta^\prime,f_\pi,f_K$ and $m_\sigma$ or $\sigma_u)$. 
However in set $d$ and $h$ instead of $m_\eta$ we have used 
$m_\sigma=680 \, \rm MeV$ for fitting the same parameters as 
$m_\sigma$ is the most sensitive to the eight-quark couplings. 

The sensitivity of sigma mass to the eight-quark coupling constant have 
been also discussed in the paper \cite{osipov2}, however instead of 
producing the eight-quark couplings, they put the values of these 
coupling constants by hand. The reason behind producing four sets
of parameters is to show the effect of different physical channels on 
the different parameters. We would like to point out here that the mass 
of the sigma meson is taken as $680 \rm \, MeV$ to fit the parameters in 
sets $b$ and $f$, due to which we obtain $m_u=12.5 \, \rm MeV$ in set $f$. 
The value of $m_u$ in set $f$ is higher than the value of 
$m_u=5-9 \, \rm MeV$ that is usually quoted in the literature. A lower mass
value of sigma meson $\sim 660 \, \rm MeV$ may give a lower value of current
u quark mass but then the constituent u quark mass will be much higher 
than the value obtained for other input parameter sets. In our analysis
we will consider the plots for the input parameter sets in the following
combinations:

\begin{itemize}
\item
Set 1 : Sets $(a,e)$ (six-quark,eight-quark). Here $m_u=5.5 \, \rm MeV$ is
held fixed. 
\item
Set 2 : Sets $(b,f)$ (six-quark,eight-quark). Here $m_{\sigma} = 680 \, \rm 
MeV$ is used to fit parameters.
\item
Set 3 : Sets $(c,g)$ (six-quark,eight-quark). Here $\sigma_u$ is used 
instead of $m_{\sigma}$.
\item
Set 4 : Sets $(d,h)$ (six-quark,eight-quark). Here $m_u=5.5 \, \rm MeV$ is
held fixed and $m_{\sigma} = 680\, \rm MeV$ is used to fit parameters.
\end{itemize}

\noindent
The resulting model parameter values of the NJL part at zero temperature,
obtained in our fit is given in the columns 2-8 of table \ref{table2}.

For the Polyakov loop part we have to consider the finite temperature 
behavior of the PNJL model. As discussed in \cite{ray} we are able to
tune the dimensionless coupling $\kappa$ in the Vandermonde term and
obtain reasonable behavior of the mean fields. We thus choose the
following set of parameters,
\begin {align}
       a_0=6.75,    a_1=-1.95,  a_2=2.625
      a_3=-7.44,   b_3=0.75,  b_4=7.5,  T_0=190 \, {\rm MeV}\nonumber
\end {align}
The remaining parameter is the dimensionless coupling $\kappa$ in the 
Vandermonde term which has been tabulated in column 9 of table 
\ref{table2}. This has been obtained by choosing suitable value of
$\kappa$ so that the pressure in PNJL model follows the Lattice QCD
data as closely as possible.

Our job is now to estimate the transition temperatures at different 
quark chemical potentials. In order to obtain this we search for the minimum 
of the thermodynamic potential which gives the temperature and density 
dependence of the fields. The point of inflection or the gap, as 
the case may be, of these fields give the transition temperature and 
chemical potential. This tusk has been carried out for both NJL 
and PNJL models.  The transition temperatures at zero chemical 
potential have been tabulated in the last
two columns of table \ref{table2}. The most important observation in this 
regard is that the introduction of the eight quark interaction lowers 
the transition temperature to the range $150-190 \, \rm MeV$. 

\begin{figure}[t]
\centering
\includegraphics[width=2.in,angle=270]{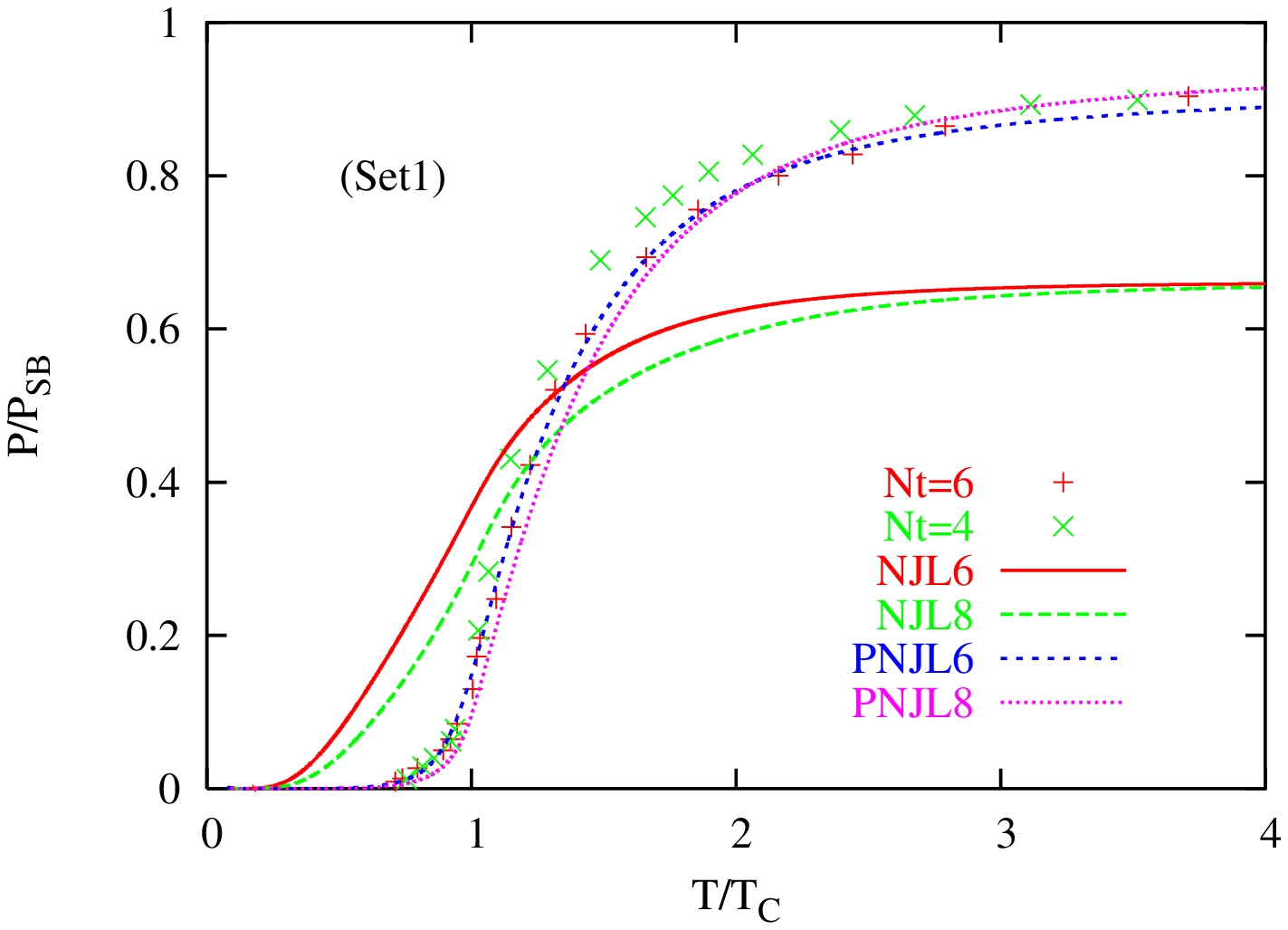}
\includegraphics[height=2.816in,width=2.in,angle=270]{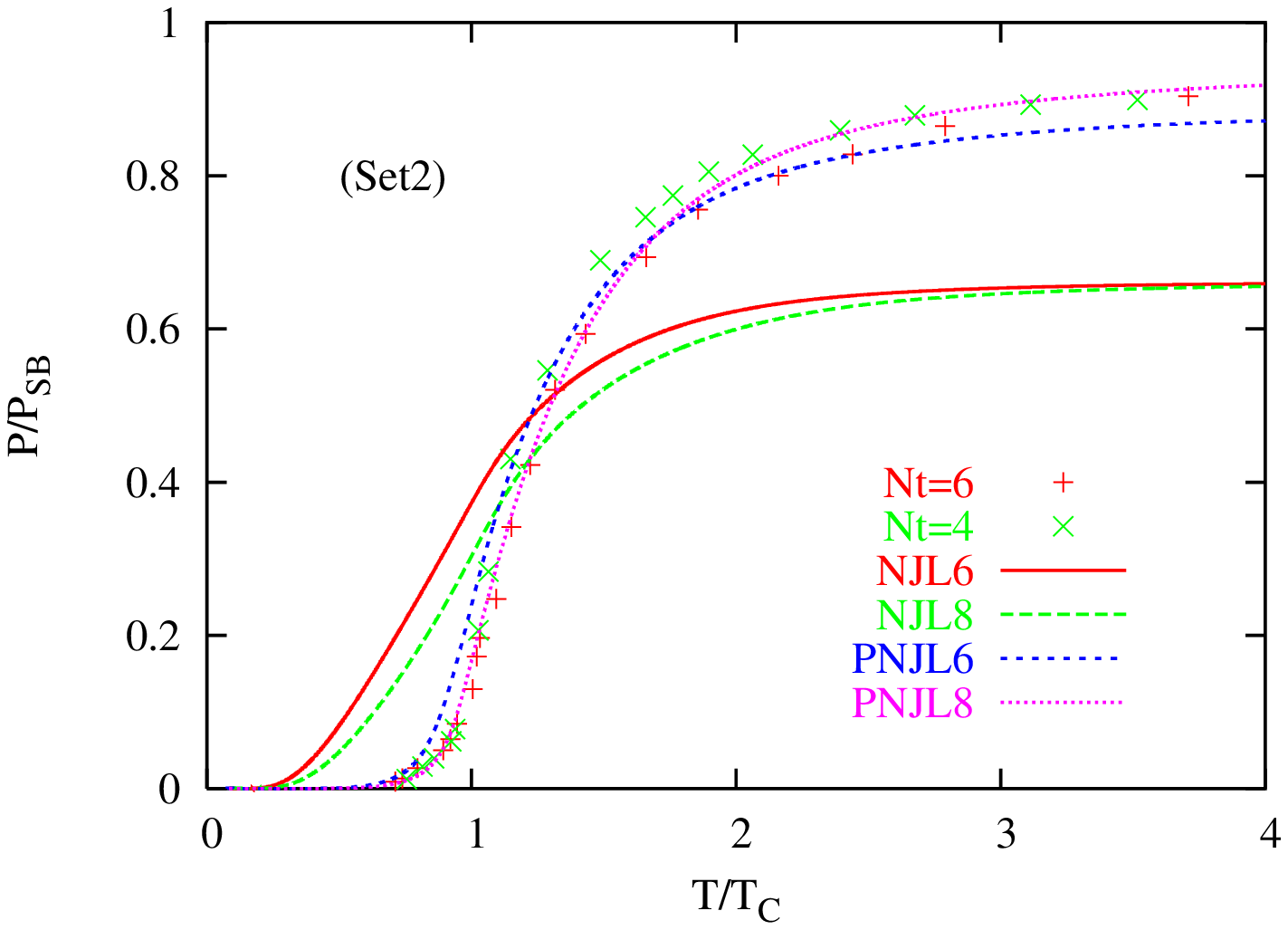}
\includegraphics[totalheight=2.8195in,width=2.in,angle=270]{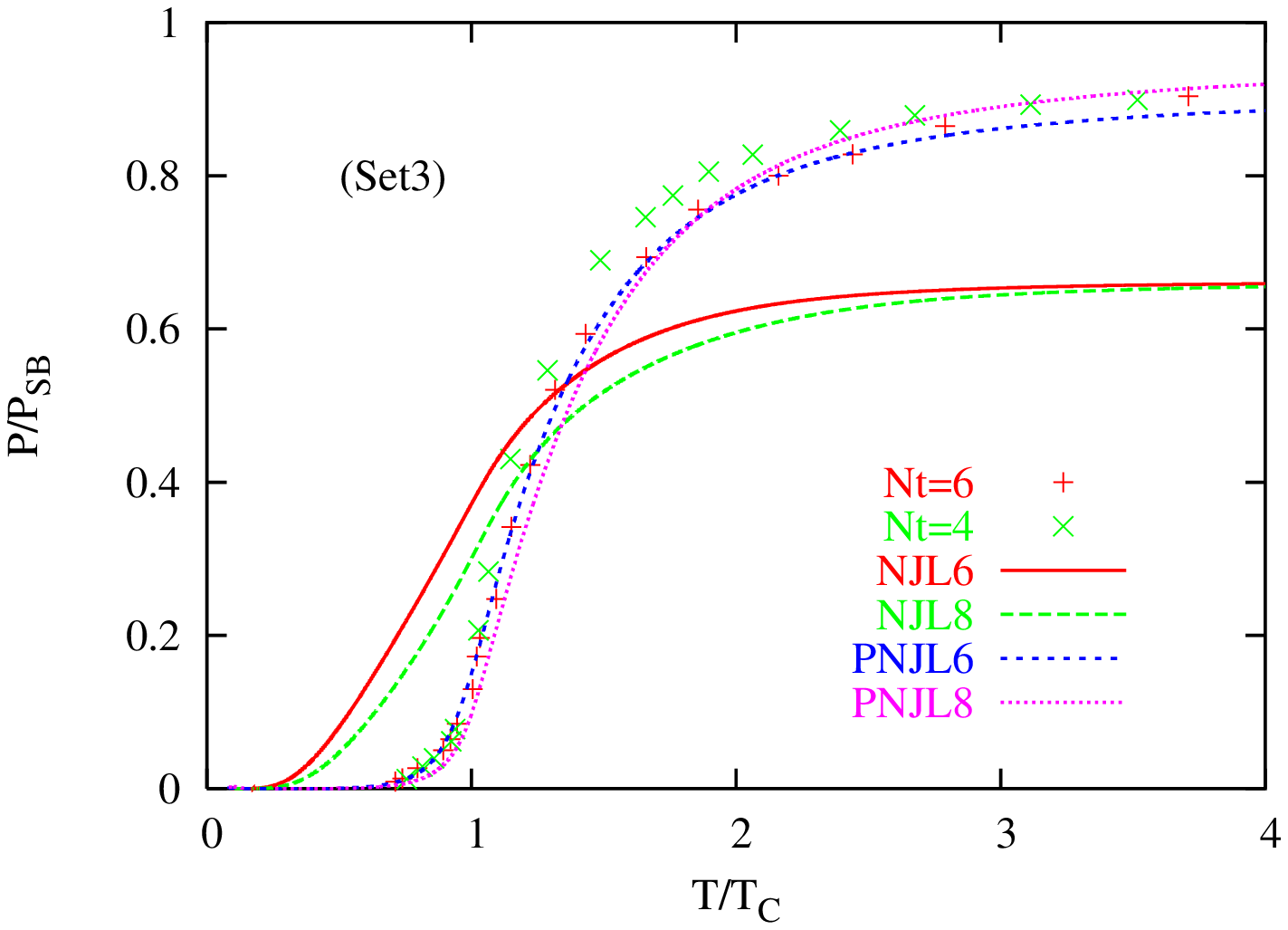}
\includegraphics[scale=2.28,width=2.in,angle=270]{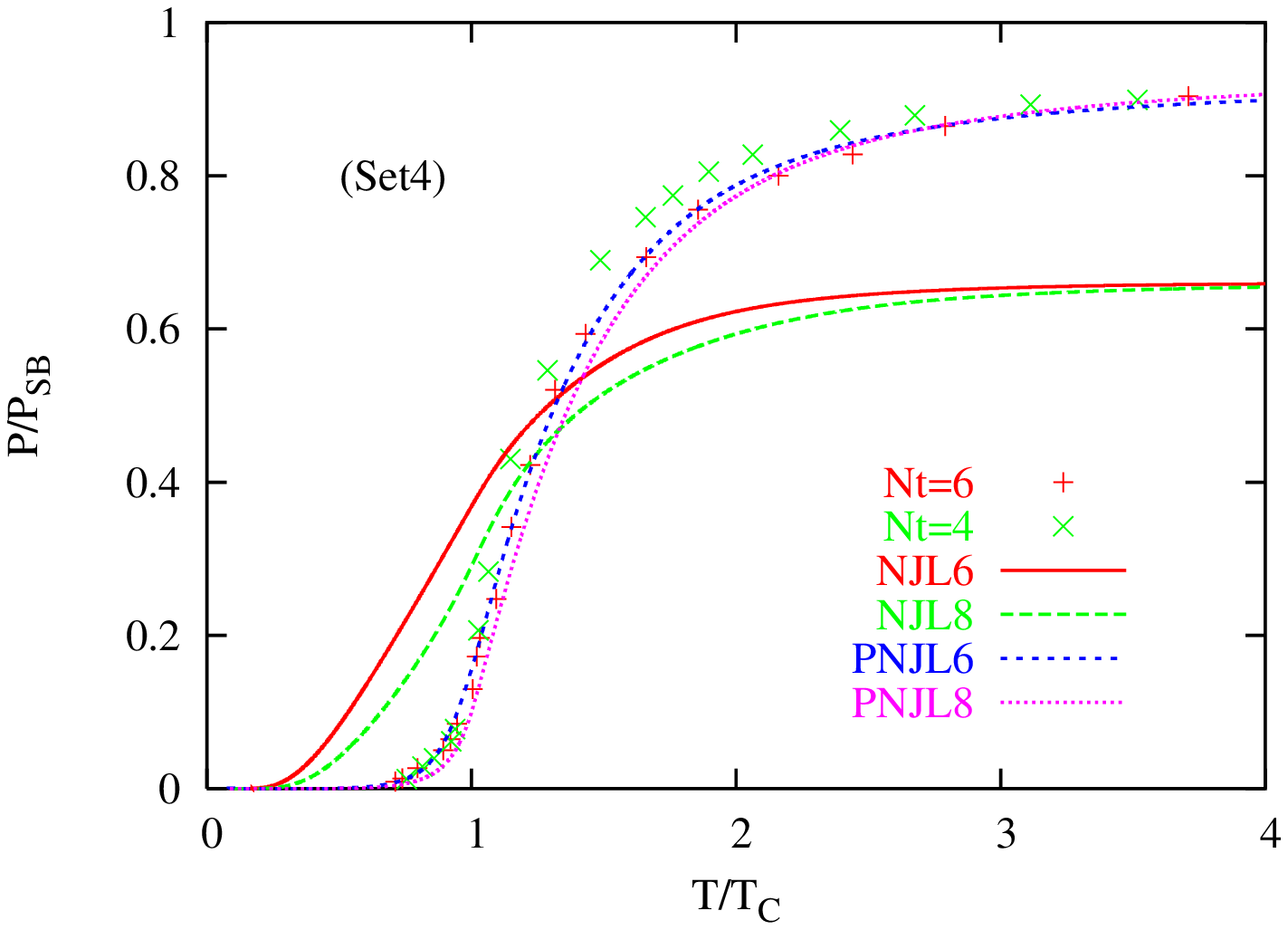}
\caption{(Color online) Variation of $P/P_{SB}$ with $T/T_C$, at $\mu=0$ for both NJL
and PNJL model. For details of the sets see text in section \ref{prmtfit}.} 
\label {prslat}
\end{figure}

The pressure of the strongly interacting matter obtained from our 
expression of the thermodynamic potential is, 
\begin {equation}
P(T,\mu)=-\Omega (T,\mu),
\label{pres}
\end {equation}
where $T$ is the temperature and $\mu$ is the quark chemical potential.
In fig. \ref{prslat} we show the variation of the scaled pressure 
$P/P_{SB}$ with $T/T_C$ in both the NJL and PNJL models for up to 
six-quark as well as up to eight-quark interactions. The plots show 
that the pressure is slightly smaller near the transition region due 
to eight-quark interaction term. In general,
the shift of pressure in the whole range of temperature ranges from 
$10\% - 20\%$. We have also compared our results with those of the 
lattice QCD data \cite{cheng} with temporal extent $N\tau=4$ and 
$N\tau=6$. In fact the pressure in the PNJL model was obtained to 
give the best possible fit to the lattice data with $N\tau=6$.
We find that for all input parameter sets it is possible to have an 
impressive similarity with the results of the lattice data. On the 
other hand the results for the NJL model differ considerably from 
the lattice data.

\begin{figure}[t]
\centering
\includegraphics[width=2.in,angle=270]{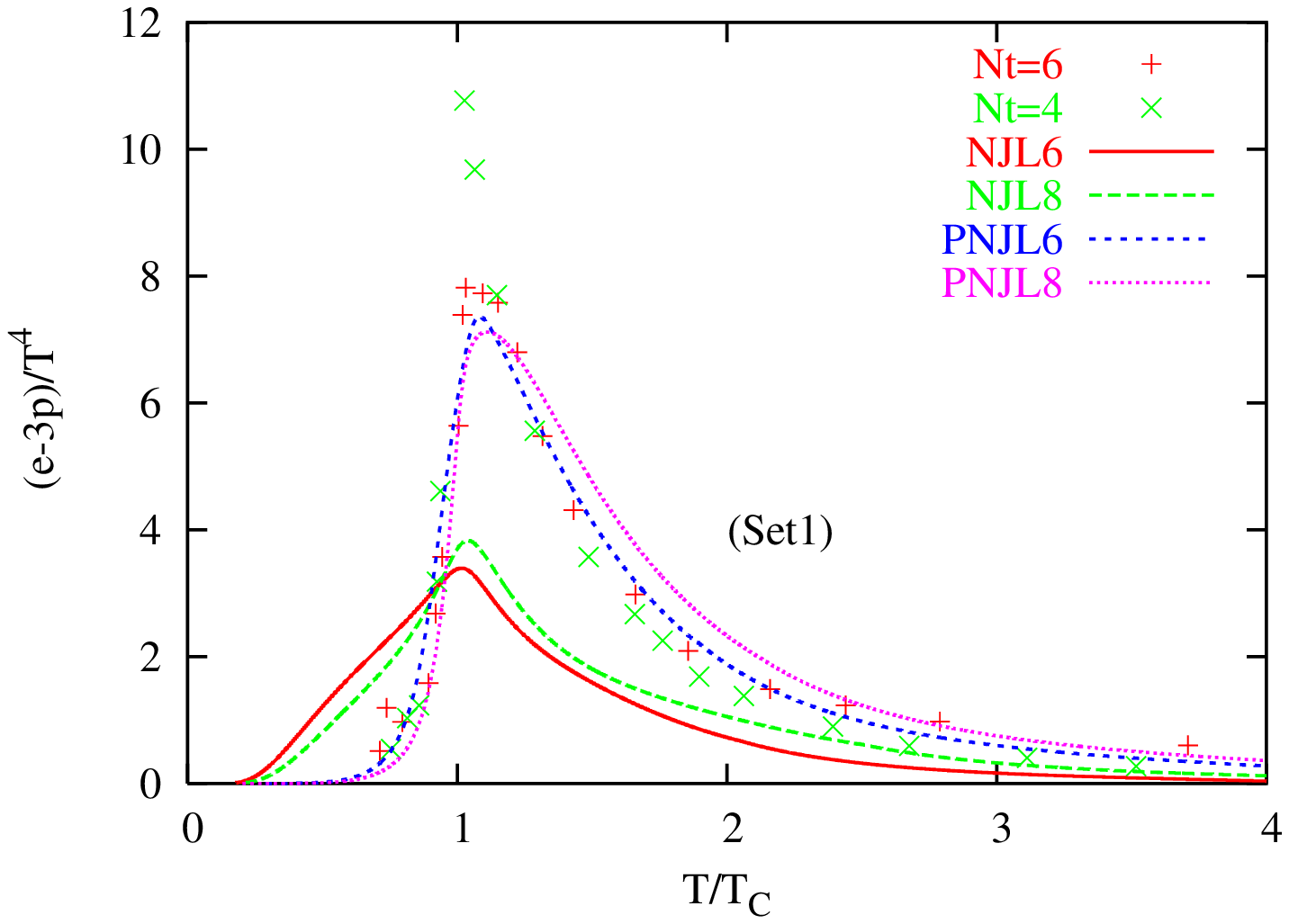}
\includegraphics[height=2.816in,width=2.in,angle=270]{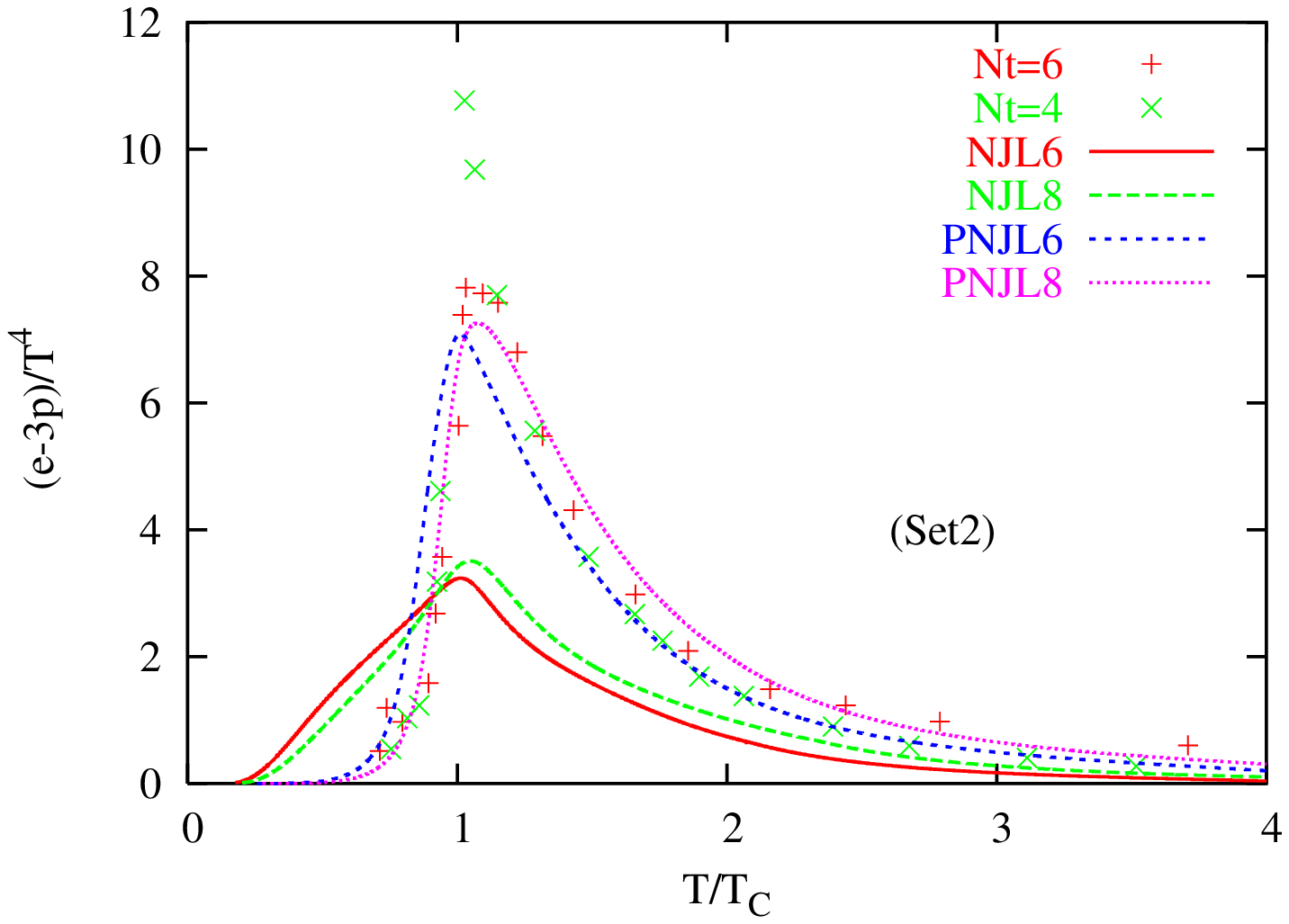}
\includegraphics[totalheight=2.8195in,width=2.in,angle=270]{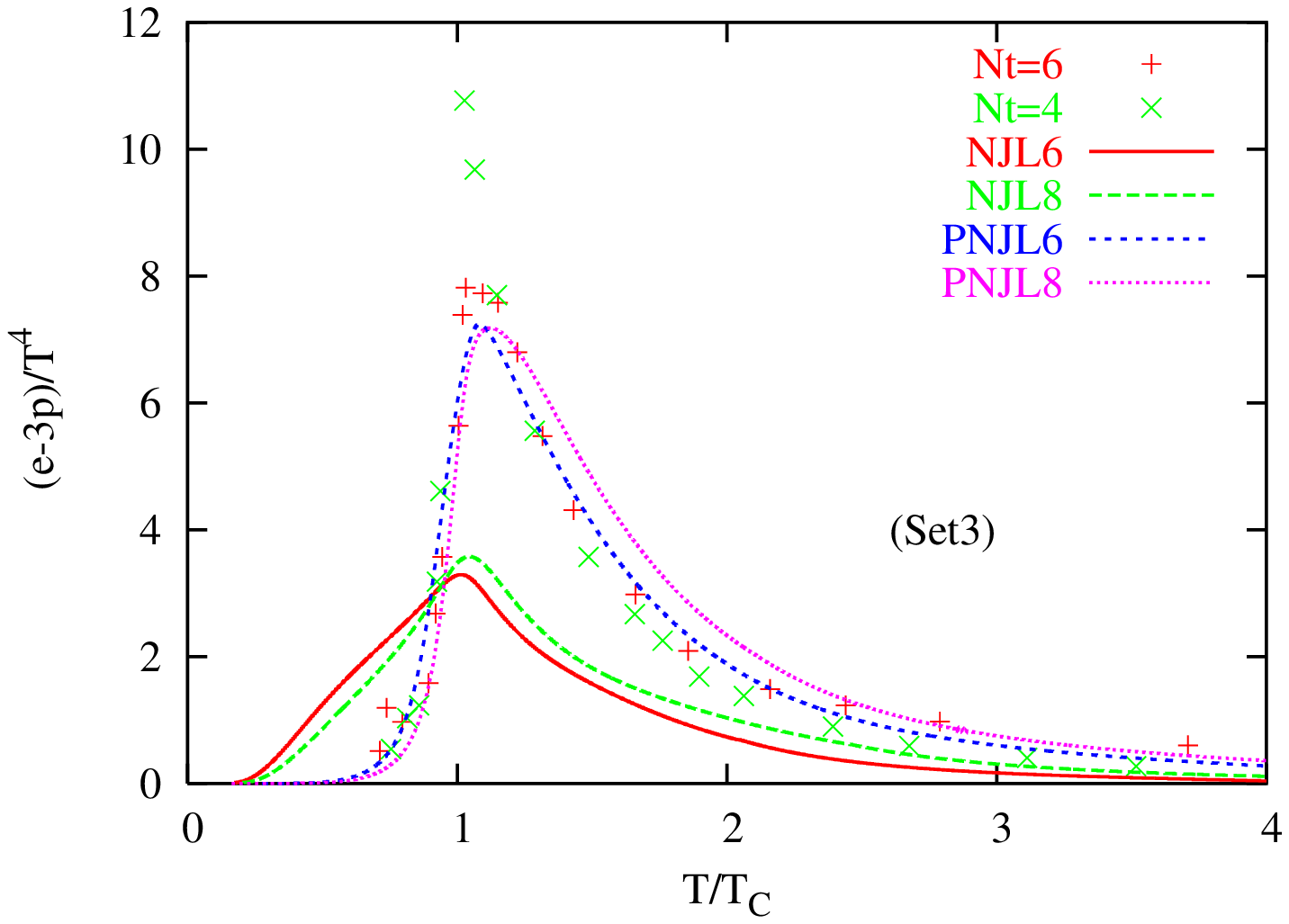}
\includegraphics[scale=2.28,width=2.in,angle=270]{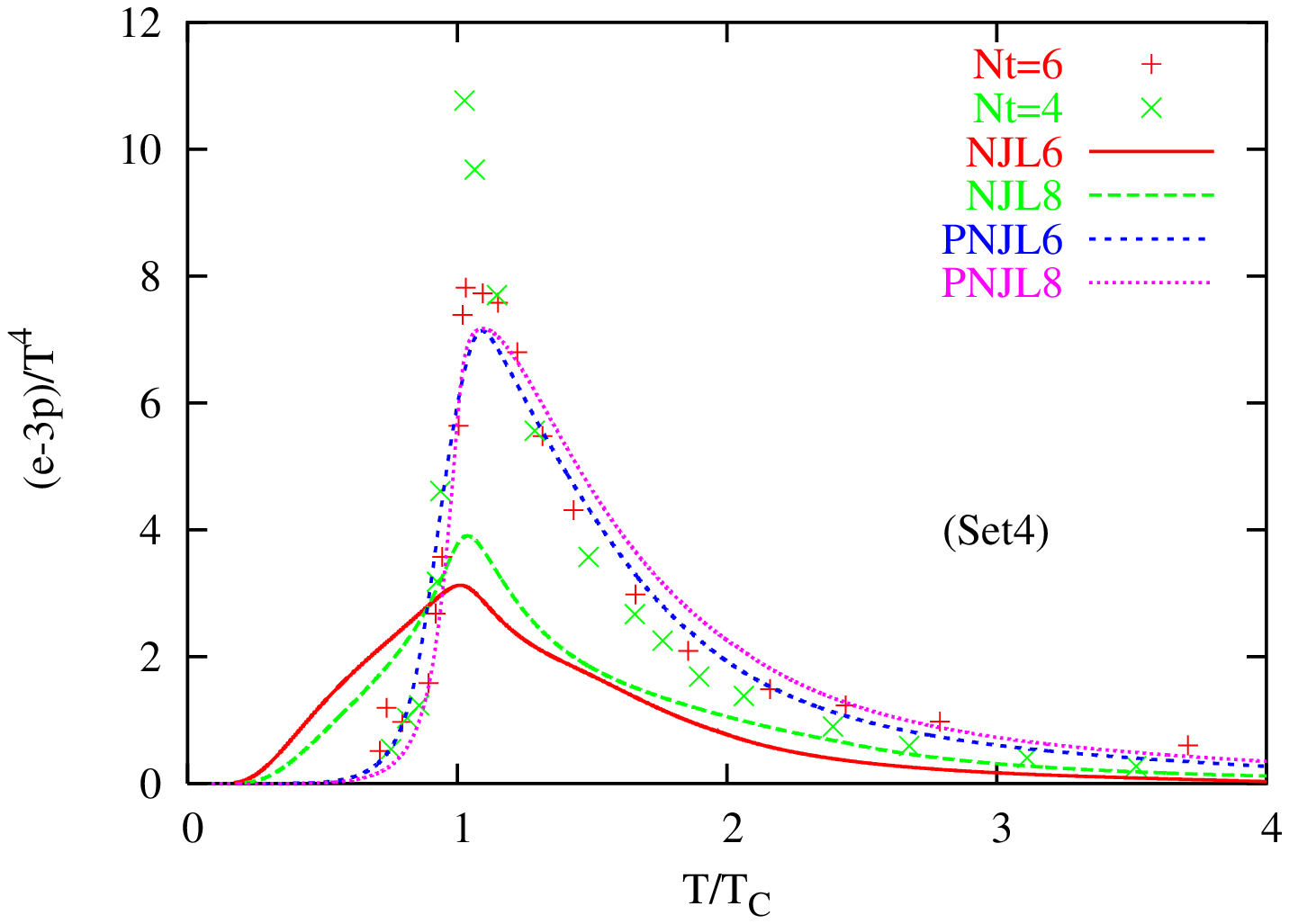}
\caption{(Color online) Variation of $(\epsilon-3p)/T^4$ with $T/T_C$, at $\mu=0$ for both NJL
and PNJL model. For details of the sets see text in section \ref{prmtfit}.} 
\label {er3plat}
\end{figure}

\begin{figure}[t]
\centering
\includegraphics[width=2.in,angle=270]{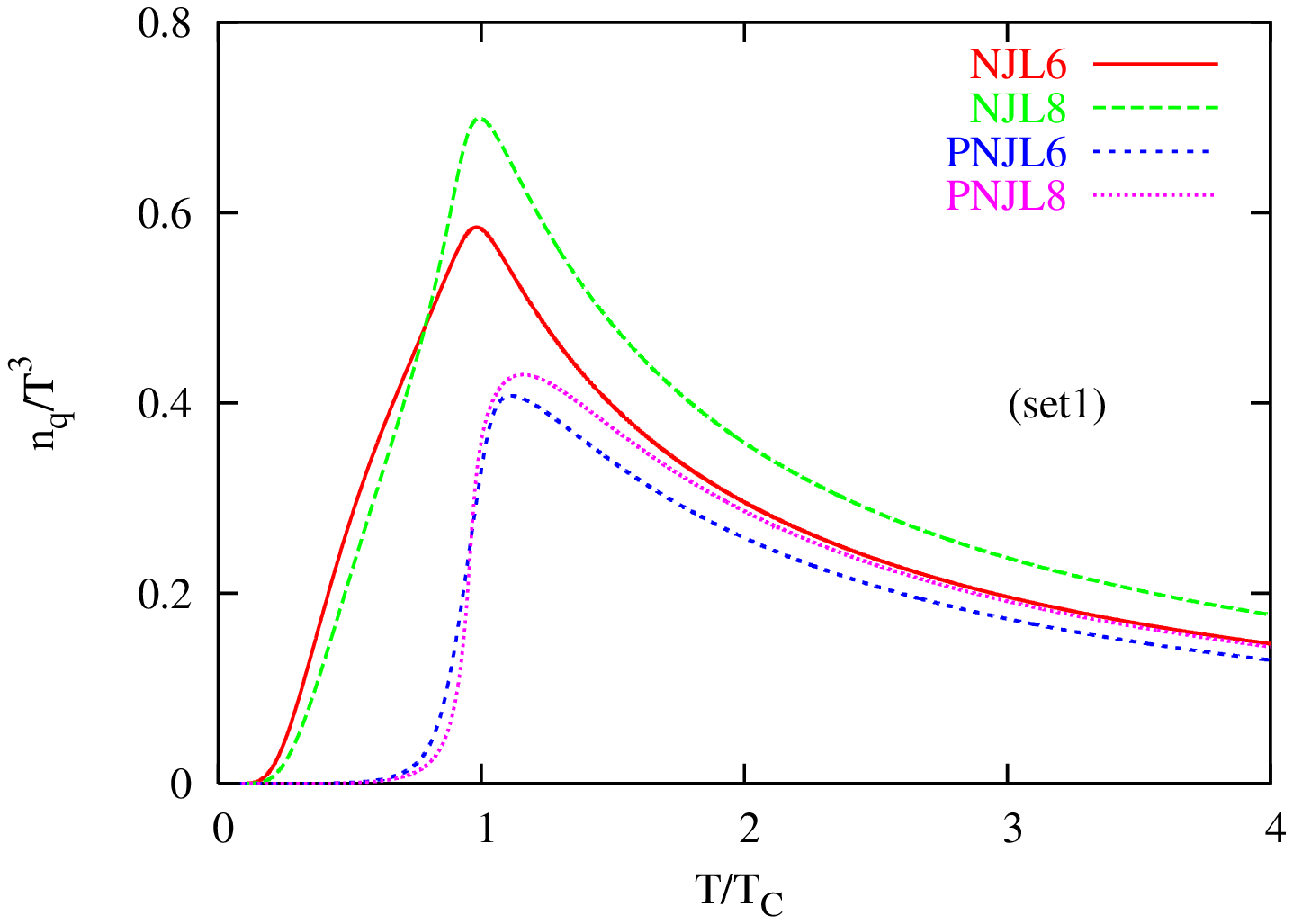}
\includegraphics[height=2.816in,width=2.in,angle=270]{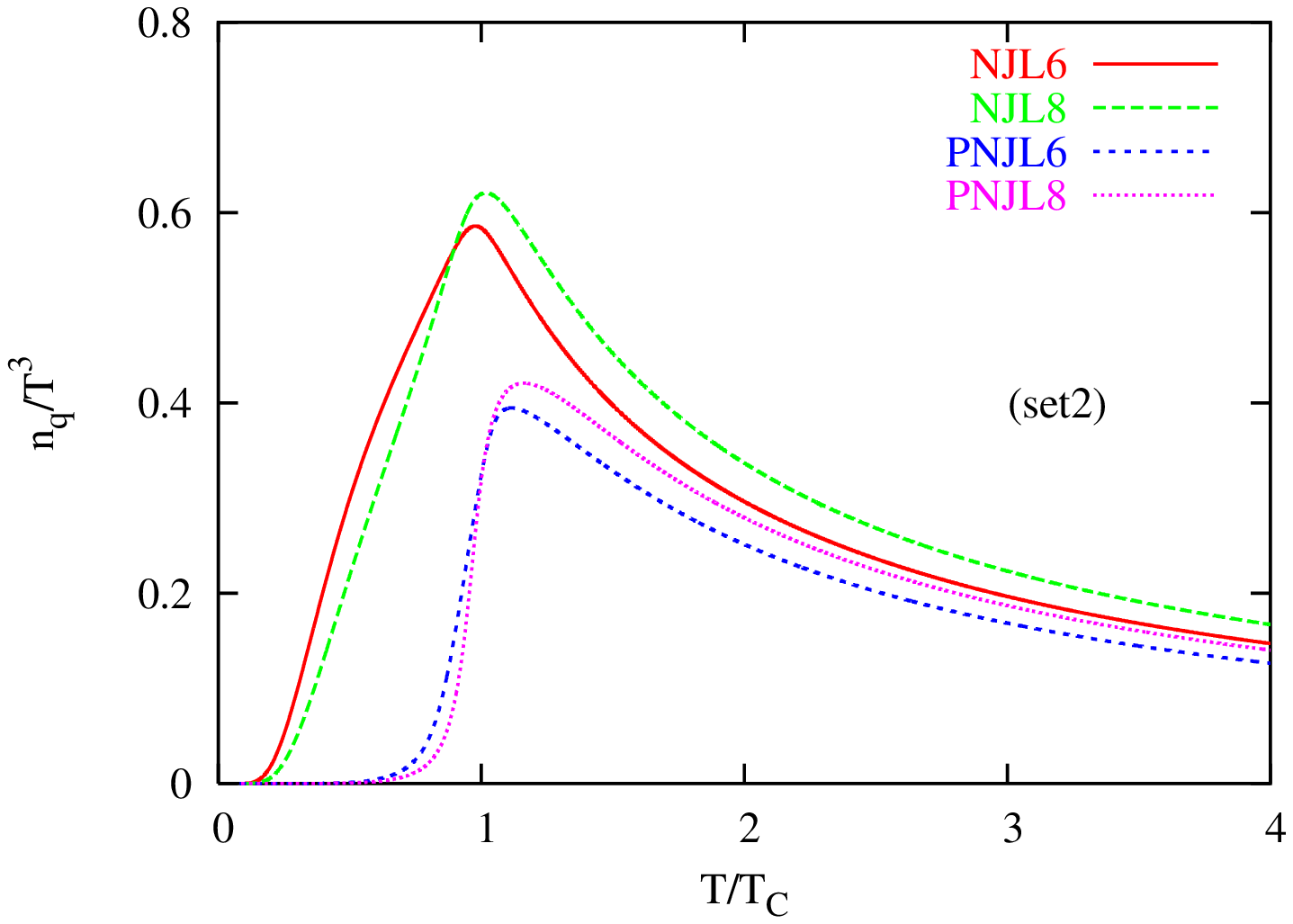}
\includegraphics[totalheight=2.8195in,width=2.in,angle=270]{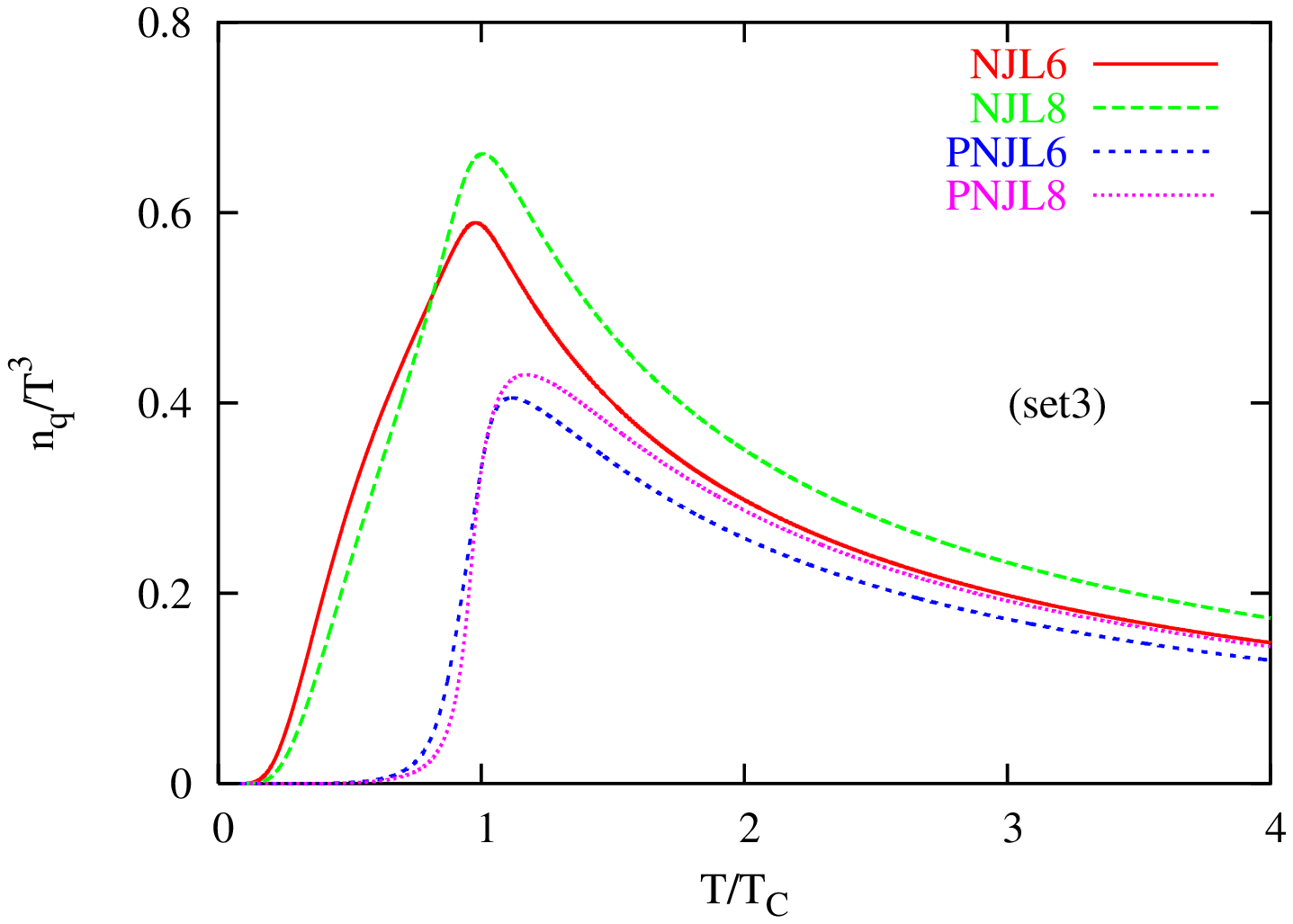}
\includegraphics[scale=2.28,width=2.in,angle=270]{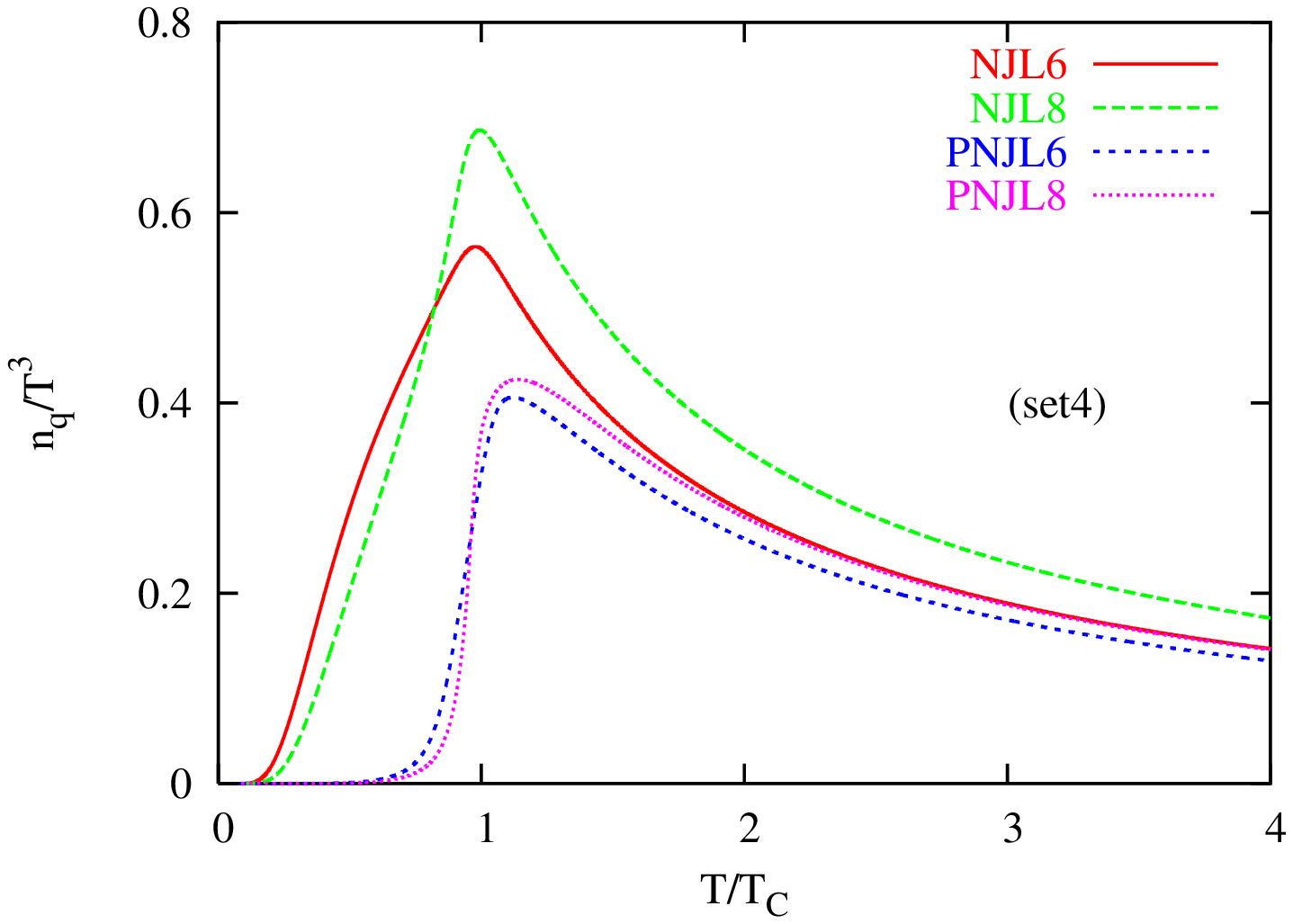}
\caption{(Color online) Variation of $n_q/T^3$ with temperature, at $\mu=100~{\rm MeV}$ 
for both NJL and PNJL models. 
For details of the sets see text in section \ref{prmtfit}.}
\label {qrkden}
\end {figure}

Another basic thermodynamic quantity 
is the energy-momentum tensor. The thermal contribution
of the trace of this energy-momentum tensor can be defined by the
difference of $\epsilon$ and $3p$, where $\epsilon$ is the energy
density and $p$ is the pressure.
So in thermodynamics the traceless energy-momentum tensor means
$\Theta_{\mu \mu}=(\epsilon-3p)=0$. The trace of the energy momentum tensor is 
vanishing at the classical level when the theory has no mass scale. We know 
that QCD in the chiral limit is scale invariant, which means that in massless
QCD, $\Theta_{\mu \mu}$ is zero unless quantum corrections are taken into 
account. Thus in a conformally symmetric theory, for
example a theory of free massless gluons, $\Theta_{\mu \mu}=0$. Therefore
this quantity measures the breaking of conformal symmetry in the interaction
theory. It is straightforward to evaluate the $(\epsilon-3p)$ or the
interaction measure from the thermodynamic potential and pressure as follows:
\begin {align}
{(\epsilon-3p)}/T^4=T\frac {{\partial (p/T^4)}}{ \partial T}
\end{align}
In fig. \ref{er3plat} we show the variation of the ${(\epsilon-3p)}/T^4$ with
temperature for both the models with and without eight-quark interaction. 
At temperatures just above $T_C$ the plots show a 
peak due to the largest deviations of $(\epsilon-3p)$ from the 
conformal limit, $\epsilon=3p$. These peaks establish a prominent structure
of trace anomaly which is consistent with lattice calculations. The 
introduction of finite value of $g_1$ and $g_2$ slightly lowers 
the peak position
in PNJL model which ensures reduction of trace anomaly due to the eight-quark
interaction term. At high temperatures the trace anomaly dropped rapidly  
similar to that obtained from the lattice data. From the plots we can 
see that in
this temperature region
much larger than $T_C$, the trace anomaly drop less rapidly for plots with 
finite value of $g_1$ and $g_2$ than the plots with  $g_1=g_2=0$. 
In all diagrams the peak position
for $N_\tau=4$ result is much higher than the $N_\tau=6$ and the peak positions
of the ${(\epsilon-3p)}/T^4$ for the PNJL model are slightly lower than
the peak position of the lattice result for $N_\tau=6$. 
It remains to be seen if this is close to the continuum limit of lattice 
data. 
At low temperature region the plots 
for all the parameter sets coincide very well with the lattice data, but at
high temperature the parameter sets with the six-quark interaction are in
better agreement with the lattice data. However the deviation of the plots 
with the eight-quark interaction term from the lattice results are also very 
small and a continuum extrapolation can be expected to yield even a better 
agreement. In pure gauge theory, it has been noted that $(\epsilon-3p)$ is quite
significant above the deconfinement temperature \cite{engels} and $\sim T^2$
for temperatures up to a few times the deconfinement 
temperature \cite{pisarski}. A similar behavior can be found in the 
PNJL model. 

The scaled quark number density is defined as:
\begin {equation}
\frac {n_q(T,\mu) }{T^3}=-\frac {1}{T^3}\frac {\partial \Omega(T,\mu)}{\partial \mu}
\end{equation}
We have plotted the quark number density as a function of $T/T_c$ at 
$\mu=100 ~{\rm MeV}$ for both the models in fig. \ref{qrkden} 
and studied the effect of eight-quark interaction.
It can be seen that at fixed values of temperature and chemical potential
the $n_q$ in PNJL model is much lower than that in the NJL model below 
$T_c$. This is evidently an effect of confinement in the PNJL model. At 
a fixed quark 
chemical potential the quark number density for the PNJL model is almost 
vanishing below the chiral transition temperature, but rises very quickly
in the vicinity of the transition temperature. However in case of the NJL
model the rise in the quark number density starts much below the 
transition temperature. Interestingly, the addition of the eight-quark 
interaction term raises the quark number significantly above $T_c$ 
for all cases at hand.
\vskip 0.3in
{\section {Phase diagrams and Critical end point}}
\begin{figure}
\centering
\includegraphics[width=2.in,angle=270]{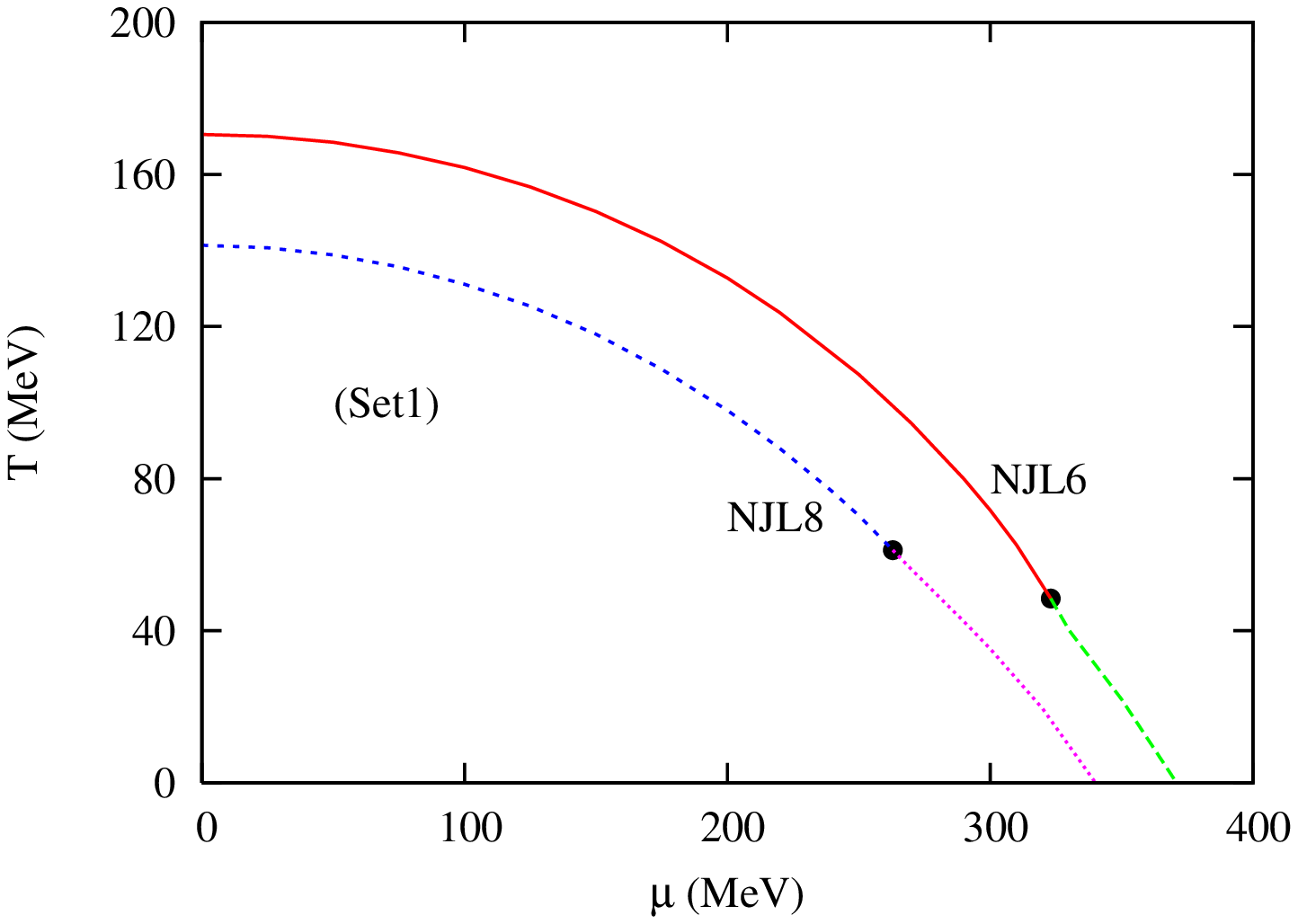}
\includegraphics[height=2.816in,width=2.in,angle=270]{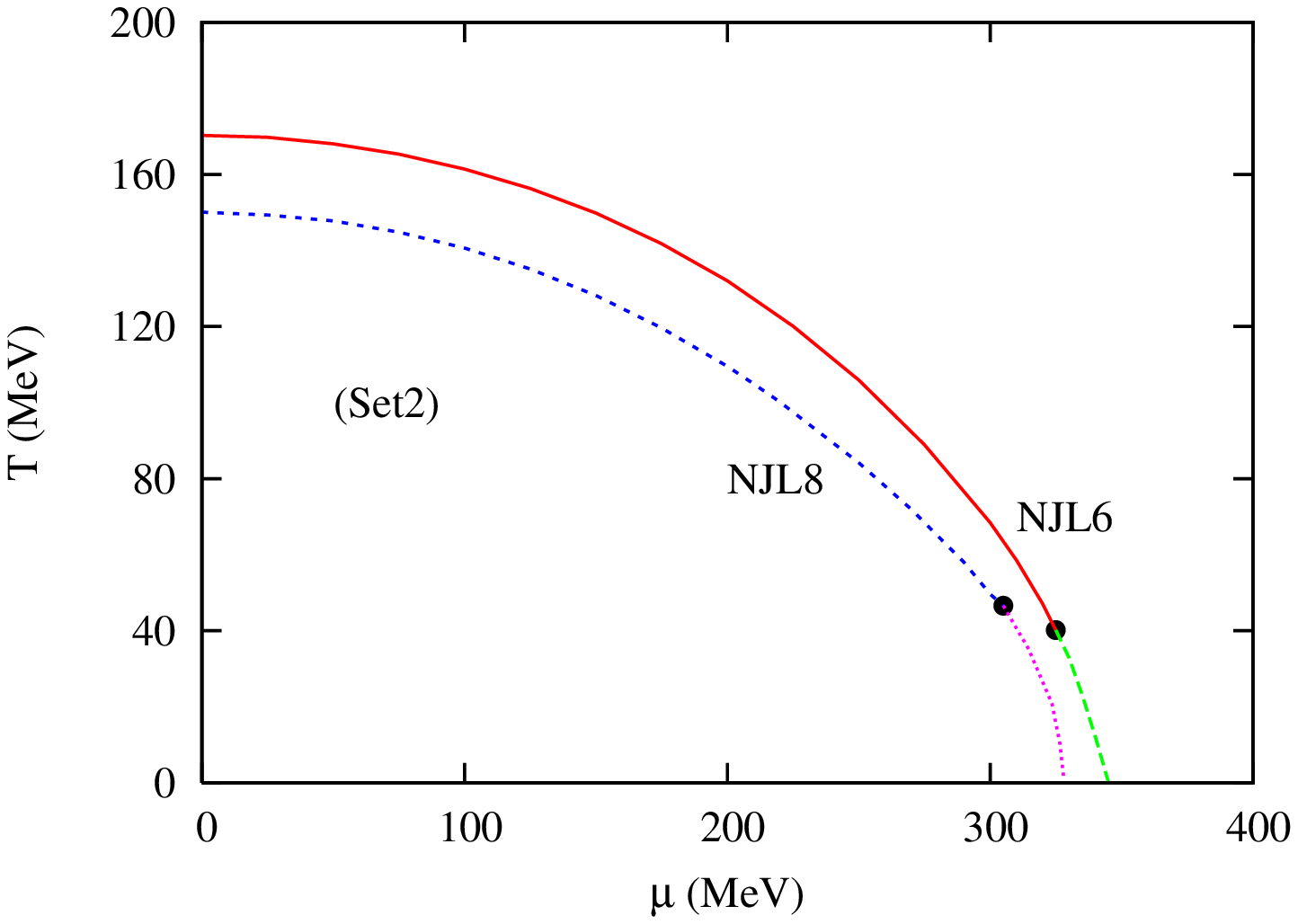}
\includegraphics[totalheight=2.8195in,width=2.in,angle=270]{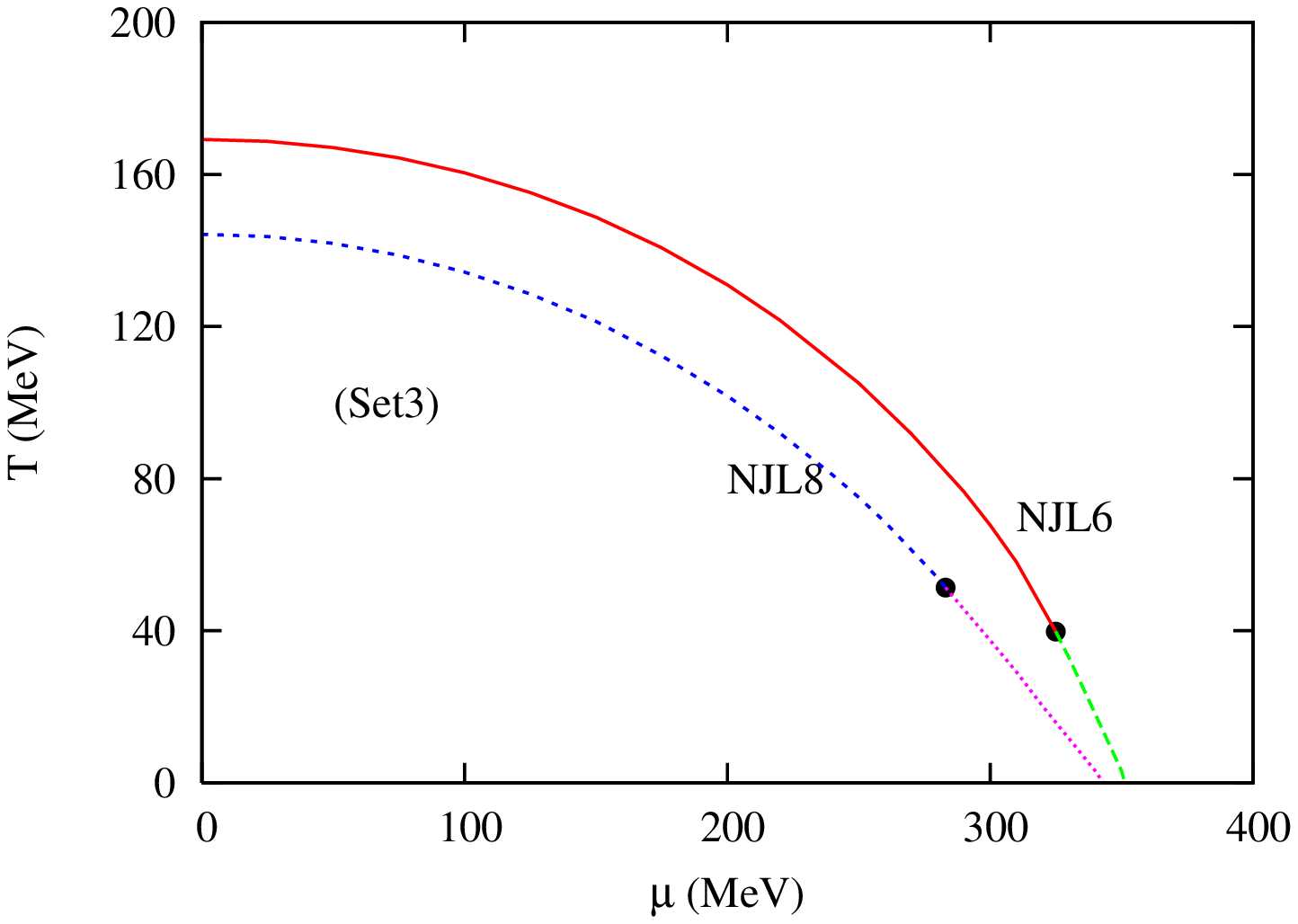}
\includegraphics[scale=2.28,width=2.in,angle=270]{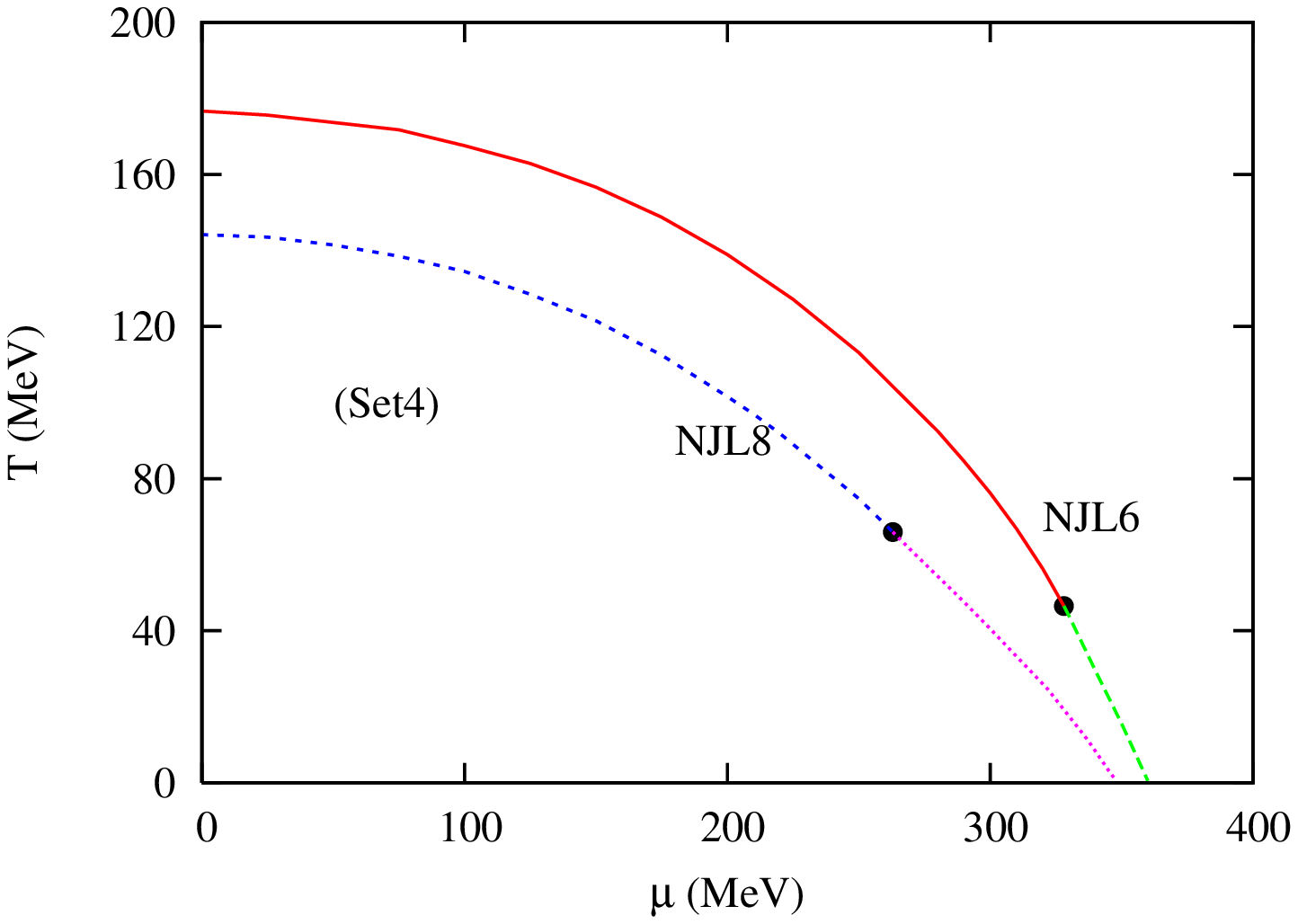}
\caption{(Color online) Phase diagram in $\mu$ with $T$ for NJL
model. For details of the sets see text in section \ref{prmtfit}.}
\label {tmnj}
\end{figure}

\begin{figure}
\centering
\includegraphics[width=2.in,angle=270]{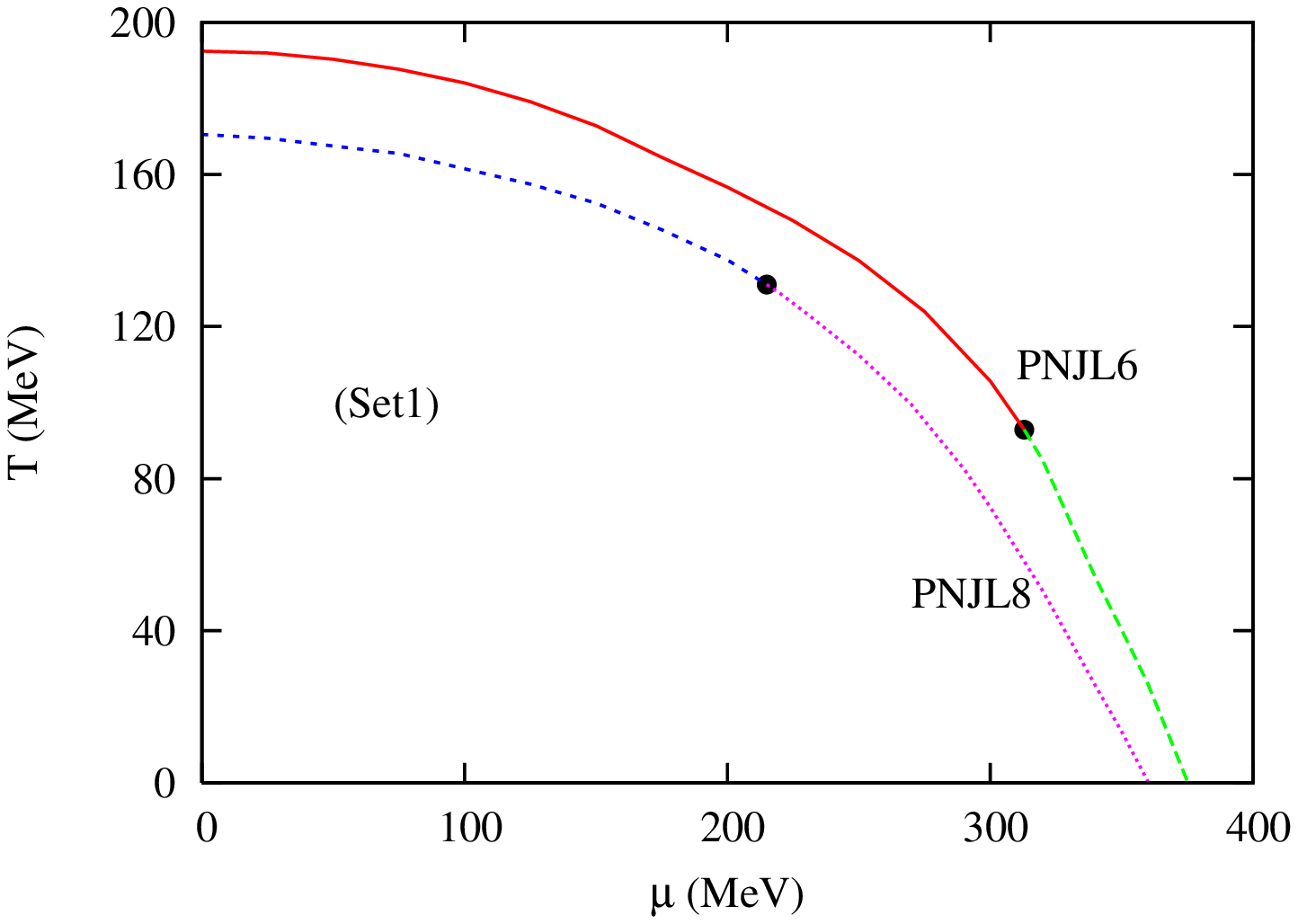}
\includegraphics[height=2.816in,width=2.in,angle=270]{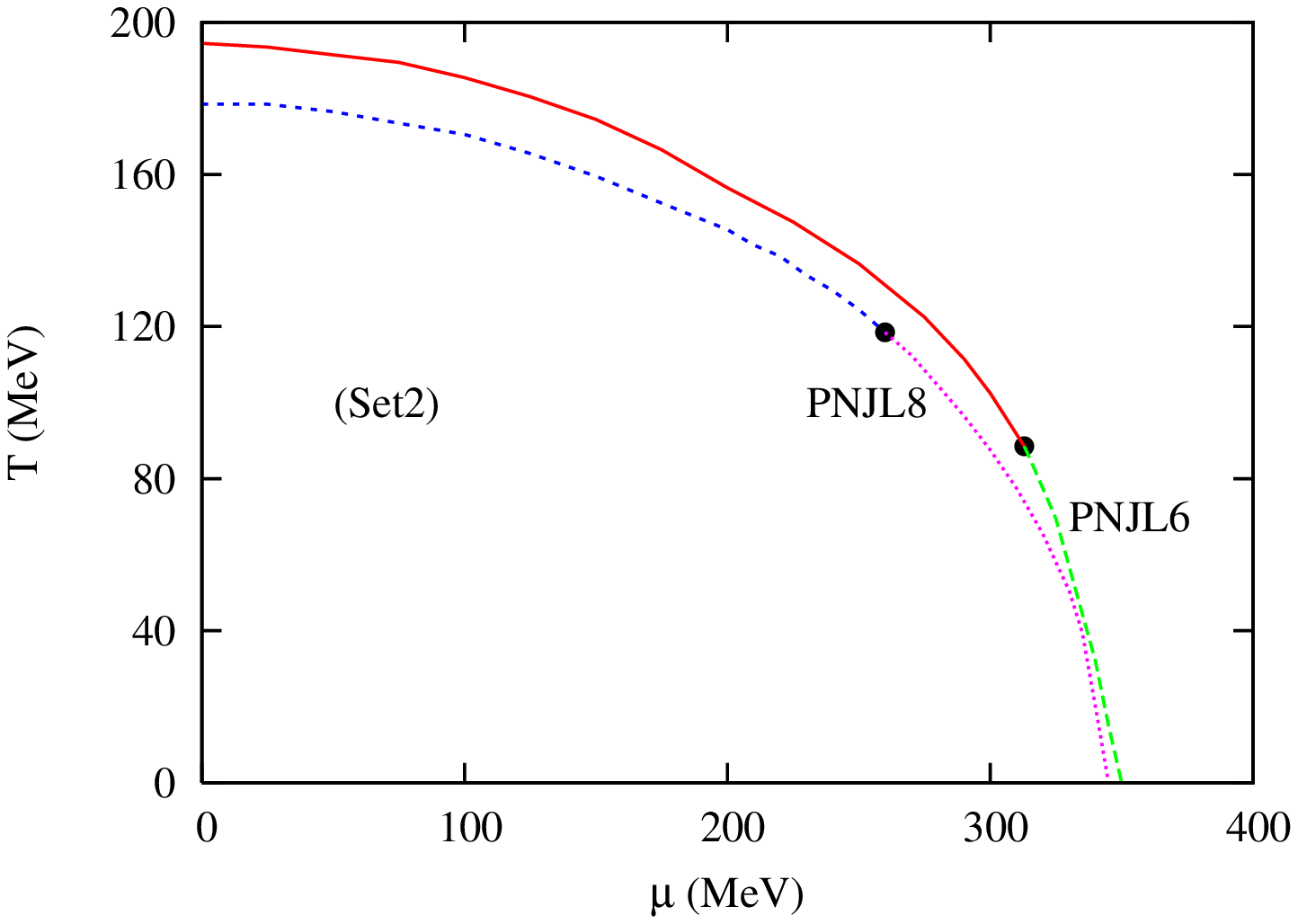}
\includegraphics[totalheight=2.8195in,width=2.in,angle=270]{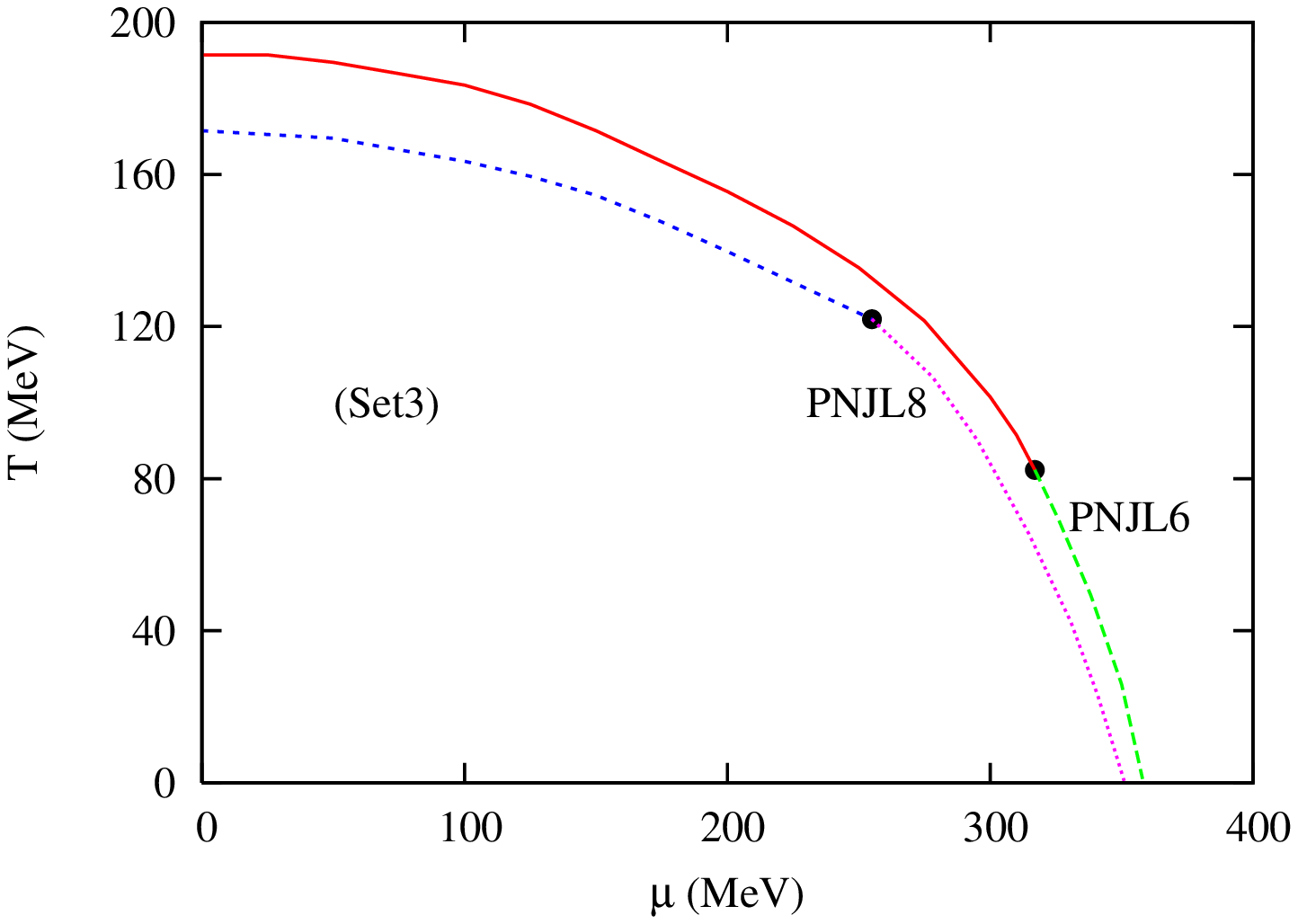}
\includegraphics[scale=2.28,width=2.in,angle=270]{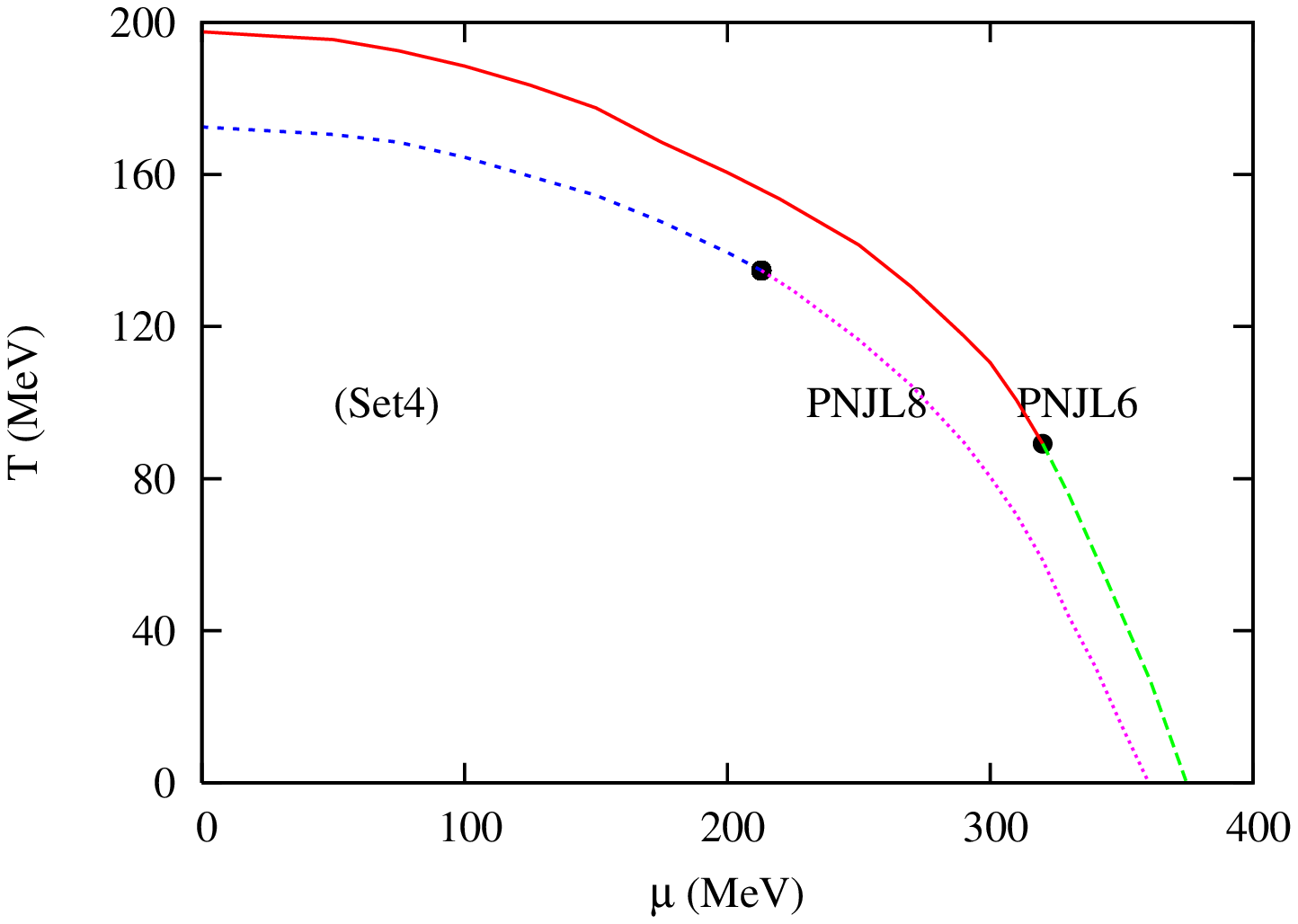}
\caption{(Color online) Phase diagram in $\mu$ with $T$ for  PNJL
model. For details of the sets see text in section \ref{prmtfit}.} 
\label {tmpnj}
\end{figure}

The position of the critical endpoint (CEP) in the phase diagram is one of 
the most interesting issue of the hot and dense strongly interacting matter.
The phase diagrams are usually obtained
by identifying the critical temperature with the temperature at which the light
quark chiral condensate has a jump (first order) or from 
the maximal point of the derivative of light quark condensate with 
respect to the temperature for different
chemical potentials. CEP is the point which separate the cross-over
transition from the first order phase transition. There have been a number of
studies in lattice QCD to find out the CEP in the $T-\mu$ diagram \cite{katz}
and also in QCD inspired models. In three flavor NJL model with the parameters
of Hatsuda-Kunihiro \cite{hat1} the location of  CEP is found to be 
$(T_C,\mu_C)=(48 \, \rm MeV, \, 324 \, \rm MeV)$, however in our 
previous work with three flavor
PNJL model we have shown the position of CEP  at
 $(T_C,\mu_C)=(92 \, \rm MeV, 314 \rm MeV)$ which establishes the fact that the location
of the CEP goes up in temperature in the PNJL model \cite{deb}. 
The reason behind this 
shift to the higher value of temperature is due to the suppression of quark
excitations at finite temperature and density by the Polyakov loop. 
However the lattice estimates of the CEP
vary cosiderably between different groups and the value of $\mu_C$ 
($\le 150 \, \rm MeV$) tend to be much lower than the PNJL value. 
   In this paper we try to 
develop a comparative study with our four input parameter sets for both NJL 
and PNJL model with and without eight-quark interaction. In fig. \ref{tmnj} and
fig. \ref{tmpnj} we have shown the phase diagrams of NJL and PNJL model with
and without eight-quark interaction term for our four input parameter sets.
In the region $T$ less than $ T_C$ and $\mu$ greater than $\mu_C $ 
the chiral and deconfinement
transitions are first order and occur almost at same $T$ and $\mu$. 
This can be realized
from the generalized Clausius-Clapeyron relation for the system with 
multiple order parameters, which shows that in case of first order phase
transition all discontinuities appear at same $T$ and $\mu$.
In table \ref{table3}
we have shown the values of CEP for different sets of parameters in both 
NJL and PNJL model. From the parameter sets we can clearly infer that 
the newly added eight-quark interaction term shifts the CEP towards 
lower $\mu$ and higher $T$ value for both NJL and PNJL model.
The recent work in lattice QCD predicts that the possible
region where CEP should exist is $\mu_C/T_C \le 2.5$ \cite{ejiri}. From the
table \ref{table3} we can observe that the PNJL model with eight-quark 
interaction term only satisfy this criteria. For the PNJL model without 
eight-quark interaction term this ratio becomes greater than 3 for all parameter
sets. Moreover in case of NJL model with eight-quark interaction term, this 
ratio is not within the allowed region. This feature establishes the 
importance of including the eight-quark interaction term in the Lagrangian.                                                                                                                              
\begin{table}
\begin{center}
\begin{tabular}{|c|c|c|c|c|}
\hline
 &\multicolumn{4}{|c|}{$(\mu_C~\rm MeV,T_c~\rm MeV) $} \\
\cline{2-5}
$\rm {Model~Index} $&set 1&set 2&set 3&set 4 \\
\hline

$\rm {NJL}~(6-quark) $&$ (323,48.5) $&$ (325,40.15) $&$ (325,39.8)
$&$(328,46.45) $\\

$ \rm {NJL}~(8-quark) $&$ (263,61.2) $&$ (305,46.55) $&$ (283,51.4)$&$
(263,65.95)$ \\
                                                                                
$\rm {PNJL}~(6-quark)$ &$ (313,92.85) $&$ (313,88.5) $&$ (317,82.25)$&$
(320,89.15)$\\
                                                                                
$\rm {PNJL}~(8-quark) $ &$(260,118.5)  $&$ (237,122.05) $&$ 
(255,121.95)$&$ (213,134.75)$\\

\hline
\end{tabular}
\caption{(Color online) The values of $(\mu_C,T_C)$ for different sets of NJL and PNJL model}
\label{table3}
\end{center}
\end{table}

\vskip 0.3in
{\section {Conclusion}}
To conclude, we have studied the bulk thermodynamic properties and 
the phase diagram of the PNJL and NJL model with eight-quark interaction.
This is the first time that 3-flavor PNJL model has been studied with 
eight-quark interaction and three-momentum cutoff. 

We have used different sets of physical observables to fix the input 
parameters for our model (including the eight quark couplings). In 
particular, we can 
observe the sensitivity of the sigma mass to the eight-quark coupling 
constants. However the value of sigma mass $m_\sigma=680 \, \rm MeV$ 
we took as one of the variable for fitting the input parameters with 
eight-quark interaction term produce
$m_u=12.5 \, \rm MeV$ which is higher than the range $5-9 \, \rm MeV$.

  The effect of eight-quark interaction on thermodynamic properties are the
main interest of our work. The variation of scaled pressure $P/P_{SB}$ with 
$T/T_C$ for four input parameters have been plotted and the results are
compared with available lattice data.  The inclusion of eight quark 
interaction decreases the pressure near the transition temperature 
and increases the pressure at higher temperature. We have also studied 
the variation of ${(\epsilon-3p)}/T^4$ with temperature. The six-quark 
interaction is already in good agreement with $N_\tau = 6$ lattice data. 
However, comparison with $N_\tau = 4$ lattice data gives us the impression 
that a continuum extrapolation may give a better match to our results 
for eight-quark interactions. 

We discuss the effect of higher order interaction term on the finite 
density results. We observe that due to the eight-quark interaction  
quark number density increases above the transition temperature.

The effect of the eight-quark interaction term on the critical end points
in the phase diagram is the main issue of our study. From 
table \ref{table3} we can
observe that the eight-quark interaction drives the CEP to a low chemical
potential and a high temperature value. Furthermore, our study 
concludes that the inclusion of
eight-quark interaction is essential to limit $\mu_C/T_C$ below $2.5$ as 
suggested by Lattice calculations.

  All these studies lead us to the conclusion that the
eight-quark interaction term in the Lagrangian has an interesting 
phenomenological implication in effective models.
\vskip 0.3in
\acknowledgments{ P.D. thanks CSIR for the financial support. 
A.B. thanks CSIR and UGC for support.
A.B. also thanks Brigitte Hiller and Saumen Datta for useful discussions.}

\end{document}